\documentclass[fleqn,usenatbib]{mnras}

\usepackage{newtxtext,newtxmath}

\usepackage[T1]{fontenc}
\usepackage{subfig}

\DeclareRobustCommand{\VAN}[3]{#2}
\let\VANthebibliography\thebibliography
\def\thebibliography{\DeclareRobustCommand{\VAN}[3]{##3}\VANthebibliography}

\usepackage{graphicx}	
\usepackage{amsmath}	

\usepackage{chemformula}

\usepackage{relsize}

\usepackage{xcolor}

\usepackage{makecell}
\newcolumntype{C}[1]{>{\centering\arraybackslash}p{#1}} 

\usepackage{nomencl} 

\usepackage[normalem]{ulem}
\usepackage{booktabs,multirow, caption}

\title[The energy distribution of the first SNe]{The energy distribution of the first supernovae}

\author[I. Koutsouridou, S. Salvadori, \'A. Sk\'ulad\'ottir, M. Rossi, I. Vanni and G. Pagnini]{
I. Koutsouridou,$^{1,2}$\thanks{E-mail: ioanna.koutsouridou@unifi.it}
S. Salvadori,$^{1,2}$
\'A. Sk\'ulad\'ottir,$^{1,2}$
M. Rossi,$^{1,2}$
I. Vanni$^{1,2}$
and
G. Pagnini$^{3}$
\\
$^{1}$Dipartimento di Fisica e Astronomia, Universit{\`a} degli Studi di Firenze, Via G. Sansone 1, 50019 Sesto Fiorentino, Italy\\
$^{2}$INAF/Osservatorio Astrofisico di Arcetri, Largo E. Fermi 5, 50125 Firenze, Italy\\
$^{3}$GEPI, Observatoire de Paris, PSL Research University, CNRS, Place Jules Janssen, 92195 Meudon, France
}

\date{Accepted XXX. Received YYY; in original form ZZZ}

\pubyear{2023}

\begin{document}
\label{firstpage}
\pagerange{\pageref{firstpage}--\pageref{lastpage}}
\maketitle
\begin{abstract}
The nature of the first Pop~III stars is still a mystery and the energy distribution of the first supernovae is completely unexplored. For the first time we account simultaneously for the unknown initial mass function (IMF), stellar mixing, and energy distribution function (EDF) of Pop~III stars in the context of a cosmological model for the formation of a MW-analogue. Our data-calibrated semi-analytic model is based on a N-body simulation and follows the formation and evolution of both Pop~III and Pop~II/I stars in their proper timescales. We discover degeneracies between the adopted Pop~III unknowns, in the predicted metallicity and carbonicity distribution functions and the fraction of C-enhanced stars. Nonetheless, we are able to provide the first available constraints on the EDF, $dN/dE_\star \propto E_{\star}^{-\alpha_e}$ with $1\leq \alpha_e \leq2.5$. In addition, the characteristic mass of the Pop~III IMF should be $m_{\rm ch}<100\:{\rm M_\odot}$, assuming a mass range consistent with hydrodynamical simulations (0.1-1000$\:{\rm M_\odot}$). Independent of the assumed Pop~III properties, we find that all $\rm[C/Fe]>+0.7$ stars (with $\rm[Fe/H]<-2.8$) have been enriched by Pop~III supernovae at a $>20\%$ level, and all $\rm[C/Fe]>+2$ stars at a $>95\%$ level. All very metal-poor stars with $\rm [C/Fe]<0$ are predicted to be predominantly enriched by Pop~III hypernovae and/or pair instabillity supernovae. To better constrain the primordial EDF, it is absolutely crucial to have a complete and accurate determination of the metallicity distribution function, and the properties of C-enhanced metal-poor stars (frequency and [C/Fe]) in the Galactic halo.

\end{abstract}

\begin{keywords}
stars: Population III -- Galaxy: formation, halo, abundances -- cosmology: first stars -- galaxies: high redshift
\end{keywords}



\section{Introduction}
\label{sec:intro}

The formation of the first stars marked a fundamental shift in the history of our universe; from a simple, homogeneous and isotropic state to the structured complexity we observe today. Their light brought an end to the so-called dark ages and initiated a period of reionization and heating of the intergalactic medium (IGM). Formed out of purely metal-free primordial gas, the first stars are commonly referred to as Population III (Pop~III) stars to distinguish them from the subsequent generations of Pop~II (or metal-poor; $Z \lesssim 0.1 \: Z_\odot$) stars and Pop~I (or metal-rich) stars. The Pop~III stars were the sources of the first metals (i.e., elements heavier than lithium) and dust grains, which they released in the surrounding medium via supernova (SN) explosions and stellar winds. 

Although critical for our understanding of the early Universe, the nature and characteristics of the first stars remain elusive. As of today, no metal-free star has been observed and theoretical models have thus far failed to reach a consensus (e.g., \citealp{Bromm2013} and \citealp{Klessen2019} for recent reviews).
Initially, first stars were thought to be very massive ($\sim100-1000\:{\rm M_\odot}$), owing to the lack of metals and dust grains, which are more efficient coolants than molecular hydrogen\footnote{Under certain circumstances, such as the suppression of $H_2$ cooling by an external radiation field \citep{Latif2013,Agarwal2016}, the critical mass for collapse can increase even further resulting into  
supermassive stars ($m \gtrsim 10^6\:{\rm M_\odot}$), which could in turn be
the progenitors of supermassive black holes, powering the first quasars \citep{Mortlock2011,Banados2018}.} and thus facilitate the fragmentation of gas clouds
 \citep{Omukai1998, Omukai2001, Abel2002, Bromm2002, OShea2007}.
Later, more detailed calculations showed that Pop~III stars can have lower masses, of the order of some tens of solar. After the initial spherically distributed gas infall, material falling with non-negligible angular momentum starts building up a rotationally supported disc around the central proto-stellar core. Radiative feedback from the nascent proto-star plays a key role in regulating this accretion process, being able to clear out the accretion disk when the star reaches a mass of $\approx 30-40\,{\rm M_\odot}$ \citep{McKeeTan2008,hosokawa2011protostellar}. Cosmological simulations that include radiative feedback confirm this picture and show that the mass spectrum of Pop~III stars is broader ($\sim 10-1000\:{\rm M_\odot}$) than previously thought \citep{hirano2014one,Hirano2015}. 
Furthermore, detailed 3D simulations investigating the formation of the first proto-stars, show that the accretion disc can be highly susceptible to fragmentation, leading to the formation of sub-solar fragments (see, e.g., \citealt{Machida2008, Clark2011, Greif2011, Dopcke2013, Stacy2016, Wollenberg2020}). However, it is still not settled whether most of these fragments migrate inwards and merge together, or are expelled from the system to survive as low mass stars \citep{Hosokawa2016, Hirano2017}. Ultimately, although the mass range of Pop~III stars is still largely unknown, it is probably biased towards massive stars, and most likely extremely broad, from $\approx 1000\,{\rm M_\odot}$ down to $<1\:{\rm M_\odot}$.

At the higher mass end, the evolution of Pop~III stars is fairly well understood. In the absence of rotation, stars with initial masses $\gtrsim 260\:{\rm M_\odot}$ collapse directly into black holes swallowing all their heavy element production \citep{Fryer2001}. Between $\sim 140$ and $\sim 260\:{\rm M_\odot}$, stars explode as pair instability supernovae (PISNe), with explosion energies and ejecta depending on their initial mass \citep{Heger2002,Takahashi2018}. The mechanism involves electron-positron pair production, which lowers the internal radiation pressure supporting the star against gravitational collapse. This pressure drop leads to a rapid contraction, which in turn ignites a runaway thermonuclear explosion that blows apart the star completely, leaving no stellar remnant behind \citep[e.g.,][]{Barkat1967, Bond1984}. At lower masses, but higher than $\sim100 \:{\rm M_\odot}$, pair-instability still takes place but the net reduction in  pressure is not sufficient to lead to the complete disruption of the star. Instead, after a series of pulsations the star produces a large iron core that likely collapses to a black hole, sweeping most heavy elements inside \citep{Heger2002}.

Contrary to PISNe, the explosion of stars with initial masses $m_\star\,\sim10-100\:{\rm M_\odot}$ leaves behind a stellar remnant and thus the final ejecta can vary strongly according to the amount of mixing and fallback\footnote{Fallback refers to the material that collapses to form a neutron star when $m_\star \lesssim 20-30\:{\rm M_\odot}$, or a black hole when $m_\star \gtrsim 20-30\:{\rm M_\odot}$ \citep{Colgate1966, Heger2010, Chan2018}.}, which depend on many factors, including the explosion energy and rotation \citep[e.g.,][]{Nomoto2006, Heger2010, Nomoto2013}. Therefore, even if the mass distribution of Pop~III stars were to be established theoretically, it is not yet settled whether these additional parameters are mass-dependent or follow separate distributions.

Currently there is no `standard model' that describes the properties of Pop~III stars. However, one can infer indirect constraints from stellar archaeology, i.e.,  the study of the oldest, most metal-poor stars in the Milky Way and its dwarf satellites (e.g.,  \citealp{frebel2015near}). These long-lived stars preserve in their atmospheres the chemical abundance patterns of their birth gas clouds, which were polluted by their ancestors: the first stars. One of the most interesting populations among them are the carbon-enhanced metal-poor (CEMP) stars, which are characterized by high relative carbon abundances [C/Fe]$\geq +0.7$ \citep[e.g., ][]{Aoki2007}. The CEMP stars are commonly divided into two major sub-classes \citep{Aoki2002, Beers2005}: 1)~those showing an excess of heavy elements produced by slow neutron-capture processes (CEMP-s); and 2)~those showing no such enhancement (CEMP-no stars). The two sub-classes are linked to different formation scenarios. CEMP-s stars represent $\gtrsim 50\%$ of all CEMP stars but are extremely rare at [Fe/H]$<-3$ \citep{Norris2013, yoon2016observational}. Their abundance pattern and the fact that $\gtrsim 80\%$ of them are members of binary systems \citep{Starkenburg_2014, Hansen2016b} are consistent with them being C-enhanced by mass transfer from an asymptotic giant branch (AGB) companion star (e.g.,  \citealp{Aoki2007, abate2015carbon}). On the contrary, CEMP-no stars are dominant among the most iron-poor stars: at least 12 out of the 14 known stars with $\rm[Fe/H]<-4.5$ are CEMP-no stars (see Section~\ref{Data} and Vanni et al. submitted). The CEMP-no stars are not preferentially found in binary systems \citep{Starkenburg_2014,Hansen2016a} or, even when they are, show no trace of mass transfer by a companion \citep{Aguado2022}, and so their chemical abundances are representative of their birth environment. 

CEMP-no stars have been observed in large numbers in the Galactic halo \citep[e.g., ][]{Yong2013b, Carollo2012, Norris2013, placco2014carbon, yoon2016observational, Bonifacio2015} and in  ultra faint dwarf (UFD) satellites (e.g.,  \citealp{Norris2010,Lai_2011,Gilmore2013, Frebel2014, Ji2016, spite2018cemp}), but also in more massive and luminous dwarf spheroidal (dSph) galaxies  (\citealt{Skul_2015}; 2023 submitted; \citealt{Susmitha2017,Chiti2018,Yoon2020}) and the Galactic bulge \citep{howes2016embla, Arentsen2021}. Their proposed zero-metallicity progenitors include: (i)~massive "spinstars" with internal mixing and mass loss \citep{Meynet2006, Maeder2015, Liu2021}, which are now strongly disfavoured by the measured high values of $\mathrm {^{12}C/^{13}C}$ in several CEMP-no stars \citep{Aguado2022,Aguado2023}; (ii)~low energy faint ($E_{51} = E_\star/10^{51} {\rm erg}<1$) or normal ($E_{51}\sim 1$) core-collapse supernovae (ccSNe) with mixing and fallback that release small amounts of iron and large amounts of carbon and other light elements (\citealp[e.g.][]{Umeda2003, Iwamoto2005, Cooke2014, Marassi2014, Tominaga2014, salvadori15}, Vanni et al. in prep). 

However, higher energy progenitors have also been found consistent with metal-poor stars. \citet{Ishigaki2014} found that the abundance pattern of the most iron-deficient star observed (SMSS J031300.36-670839.3; \citealt{Keller2014}) is well reproduced both by a $E_{51}=1$ ccSN and by a $E_{51}\geq 10$  Pop~III hypernova. \citet{Ishigaki2018} suggested that more than half of a sample of $\sim200$ extremely metal-poor (EMP; [Fe/H]<-3) literature stars are best fitted by a $m_\star = 20\:{\rm M_\odot}$ ($E_{51}=10$) hypernova model. \citet{Placco2015}, however, who analyzed a subset of these stars, found systematically lower energy-progenitors for them (see Table~2 in \citealp{Ishigaki2018}). \citet{Ezzeddine2019} argued that the CEMP-no star HE 1327-2326 shows an imprint of a $E_{51} = 5$ assymetric hypernova explosion. More recently, \citet{Skuladottir2021} and \citet{Placco2021} discovered two ultra metal poor stars ($\rm[Fe/H]<-4$) in the Sculptor dSph galaxy and in the Galactic halo, respectively, with very low [C/Fe] and abundance patterns indicating that they descend from $E_{51}=10$ Pop~III hypernovae (see also Sk\'{u}lad\'{o}ttir et al. 2023). 

Besides identifying individual Pop~III progenitors, several studies have employed galaxy formation models of MW-analogues to investigate the properties of Pop~III stars in a statistical manner, including their spatial distribution \citep[e.g.][]{white2000first, Brook2007,Tumlinson2010,Salvadori2010, Ishiyama2016, Starkenburg2017,Hartwig2022} and the impact of their initial mass function (IMF) on the abundances of metal-poor stars \citep[e.g.][]{Salvadori2007, Komiya2010, deBen2017, Hartwig2018a, Sarmento2019, Tarumi2020}. However, none of these studies has examined the full parameter space of Pop~III stars - IMF, mixing, explosion energy - and none has considered the existence of high $E_{51}$ primordial SNe.

In this work, we aim to fill this gap and explore how varying the energy distribution of the first SNe affects the properties of the stars surviving until $z=0$ in the Galactic halo. To this end, we develop a new semi-analytic model (SAM), named {\sc NEFERTITI} ({\it NEar-FiEld cosmology: Re-Tracing Invisible TImes}), that traces the formation and evolution of individual Pop~III and Pop~II/I stars,  accounting for all the unknown related to primordial star-formation. NEFERTITI can run on any merger-tree or dark matter (DM) simulation to shed light on the earliest phases of star-formation. Here, we combine it with a DM simulation of a MW analogue. This allows us to follow in detail the early chemical evolution of the first star-forming halos and link them with the observed properties of Galactic halo stars in order to constrain the nature of Pop~III stars and of the first SNe. 

\section{The model}
\label{sec:model}

This Section introduces our newly developed SAM {\sc NEFERTITI}, which in this work is coupled with a cosmological N-body simulation for the formation of a MW analogue. {\sc NEFERTITI} builds upon our previous works with the SAM {\sc GAMETE} \cite{Salvadori2010, graziani2015galaxy, graziani17, pacucci2017gravitational, Pagnini2023}, but represents significant advancements, allowing us to account for: (i) finite stellar lifetimes of both Pop~III and Pop~II/I stars, i.e., relaxing the instantaneous recycling approximation; (ii) the incomplete sampling of the IMF for both Pop~III and Pop~II/I stars; (iii) an unknown energy distribution function for Pop~III stars exploding as SNe.

 In the following, we describe briefly how the N-body simulation traces the hierarchical growth of dark-matter (DM) haloes (Section~\ref{DM}) and, in detail, how the evolution of their baryonic content is followed by the SAM (Section~\ref{baryons}). Finally, in Section~\ref{calibration} we present the calibration of the free parameters of our model.

\subsection{The \textit{N}-body simulation}
\label{DM}
We use a cold dark matter N-body simulation of a MW analogue \citep{scannapieco2006spatial,Salvadori2010} that has been carried out with the {\sc GCD+} code \citep{kawata2003gcd+} using a multi-resolution technique \citep{kawata2003multiwavelength}. The highest resolution region has a radius of four times the virial radius\footnote{For comparison, the virial radius of the MW is estimated at $R_{\rm vir, MW}= 200-290\:{\rm kpc}$ (e.g.,    \citealp{dehnen2006,posti2019}).} of the system, $R_{\rm vir}=239\:{\rm kpc}$ at $z=0$, and a softening length of $540\:$pc. The system comprises $\sim10^6$ DM particles of mass $\sim7.8 \times 10^5 \: {\rm M_\odot}$, i.e., a virial mass $M_{\rm vir}\approx7.8 \times 10^{11}\:{\rm M_\odot}$, consistent with observational estimates for the MW ($M_{\rm vir, MW} \approx 6 - 25 \times 10^{11}\:{\rm M_\odot}$; see \citealp{wang2015MWmass} and references therein). A low-resolution simulation including gas physics and star formation has been used to confirm that the initial conditions will lead to the formation of a disc galaxy. The position and velocities of all DM particles are stored, at each snapshot of the simulation, and a friend-of-friends algorithm, with a linking parameter $b = 0.15$ and a threshold number of particles of 50, is used to identify the virialized DM haloes. The timestep between the snapshots is $\Delta t_z\approx22\,$Myr at $17<z<8$ and $\approx110\,$Myr at $z<8$.

\subsection{The modelling of baryons}
The SAM follows the flow of baryons from the intergalactic medium (IGM) into the DM haloes, the formation of stars and stellar evolution within each galaxy, and the return of mass and metals into the interstellar medium (ISM) and the IGM through stellar feedback. In order to resolve the evolution of the most massive stars, we adopt a shorter sub-timestep of $\delta t_s = 1\:$Myr for the SAM. At the end of each timestep $\Delta t_{\rm z}$ of the N-body simulation, the stellar and gas mass within each halo is equally distributed among all its DM particles. The baryons then follow the course of their respective DM particles in the next integration step of the N-body simulation. This way we can extract the spatial distribution of all stellar populations throughout the Galaxy's assembly history. 

At $z<3$, when the DM halo of our central galaxy (i.e., the MW) has grown significantly, the assumption that the newly formed stars are equally distributed among its DM particles is no longer valid. Indeed, we know that a disc should form, leading to a more centrally confined star formation. However, this approximation is suitable for investigating the ancient very metal-poor stellar halo populations we are interested in, which form at $z > 5$ in smaller progenitor halos. Indeed, we have confirmed that ancient stars (born before $z \sim 5$) inhabiting our mock MW's halo at $z=0$ have been mostly acquired via mergers, i.e., they form ex-situ.

\label{baryons}
\subsubsection{Gas accretion}

According to the traditional view of galaxy formation, all infalling gas is initially shock heated to the virial temperature, $T_{\rm vir}$, of the DM halo and forms a quasi-static atmosphere, which then cools  radiatively from the inside out and falls onto the central galaxy (hot mode accretion; \citealp{Rees1977, Silk1977, White1978}). More recent simulations have revealed a new paradigm in which part of the gas enters the halo along cold, dense filaments and accretes directly onto the galaxy without being shock-heated (cold-mode accretion; \citealp{Keres2005, Dekel2006, Cattaneo2020}). The latter is the dominant accretion mode for virial masses $M_{\rm vir}<10^{11}\:{\rm M_\odot}$, and is therefore more relevant for our study that focuses on very metal-poor stars (VMP; $\rm[Fe/H]\leq-2$), which form at $z>5$ in our simulation, when the maximum halo mass is $M_{\rm vir}\sim 5 \times 10^{10}\:{\rm M_\odot}$.

In the lowest-mass halos, baryonic infall may be substantially reduced due to photodissociating and photoionizing radiation, since gas cannot cool and accrete onto halos with virial temperature, $T_{\rm vir}$, lower than that of the IGM \citep[e.g., ][]{Blanchard1992}. To account for this effect, we assume that there is no gas accretion onto halos with $T_{\rm vir}<T_{\rm SF}$, where $T_{\rm SF}=2 \times 10^3\:$K at $z>z_{\rm rei}=6$, and $T_{\rm SF}=2 \times 10^4\:$K at $z \leq z_{\rm rei}$ when the Milky Way environment is fully ionized \citep{Salvadori2014}.

For haloes with $T_{\rm vir}>T_{\rm SF}$, we assume that gas from the intergalactic medium is continuously added into their cold filaments at a rate that is proportional to their dark matter growth:
\begin{equation}
\dot{M}_{\rm fil, accr} = f_{\rm b}\dot{M}_{\rm vir},
\end{equation}
where $f_{\rm b} =  \Omega_b / \Omega_m$ is the universal baryon fraction.

The gas within the filaments is subsequently assumed to stream onto the central galaxy on a free-fall timescale:
\begin{equation}
t_{\rm ff} = \Big({\frac{3 \pi}{32 {\rm G} \rho}}\Big)^{1/2},
\label{e:tff}
\end{equation}
where ${\rm G}$ is the gravitational constant and  $\rho = \rho_{\rm 200}(z)$ is the total (dark+baryonic) mass density of the halo at redshift $z$. Hence, the gas accretion rate onto a galaxy and the variation of its filaments' gas mass are given, respectively, by  
\begin{equation}
\dot{M}_{\rm gas, accr} = \frac{M_{\rm fil}}{t_{\rm ff}}
\label{e:gas_accr}
\end{equation}
and
\begin{equation}
\dot{M}_{\rm fil}  = \dot{M}_{\rm fil,accr} - \dot{M}_{\rm gas, accr}.
\end{equation}

\subsubsection{Star formation}
Star formation (SF) occurs in a single burst at each sub-timestep $\delta t_s$, at a rate given by the cold gas mass, $M_{\rm gas}$, within the galaxy, the free-fall time (Eq.~\ref{e:tff}), and the SF efficiency  $\epsilon_{\rm SF}$, which is a free parameter of our model: 
\begin{equation}
{\rm SFR} = \epsilon_{\rm SF} \frac{M_{\rm gas}}{t_{\rm ff}}.
\label{e:sfr}
\end{equation} 
Following \citet{salvadori15}, we assume that the SF efficiency in minihaloes with $T_{\rm vir} \leq 2 \times 10^4\:$K is reduced by a factor of $\epsilon_{\rm H_2}=2\epsilon_* [ 1 + (2 \times 10^4\,{\rm K}/T_{\rm vir})^3]^{-1}$, to account for the ineffective cooling by molecular hydrogen \citep{salvadori2012}.

At each timestep $\delta t_s$, and for each halo, we compute the stellar mass formed, $M_*={\rm SFR} \cdot \delta t_s$, and form a Simple Stellar Population (SSP) only if $M_*$ is greater or equal to the maximum stellar mass $m^{\rm max}_{\star}$ allowed by the IMF\footnote{Throughout the text, the index ($*$) is used to refer to the total stellar mass of a galaxy, while the index ($\star$) is used to refer to the mass of individual stars.}. This way we ensure that stars throughout the whole mass range of the assumed IMF can be represented (see Section~\ref{sec:stellev}). Each SSP is characterized by its formation time $t_{\rm form}$, the number and initial masses of its stars and their elemental abundances (which are equal to the ones of the gas in their host halo at $t_{\rm form}$). 

Our star formation model is calibrated to act on the total gas content within a galaxy (see Section~\ref{calibration}); we do not differentiate here between the different phases (e.g.,  cold, warm, molecular, atomic) of the ISM. In addition, our model ignores some physical mechanisms, such as merger/instability induced starbursts, mass quenching (including AGN and halo quenching\footnote{AGN quenching refers to feedback from an active galactic nucleus (AGN) that can eject cold gas from within a galaxy and/or prevent the hot gas
surrounding it from cooling. Halo quenching refers to the disruption of cold filamentary flows by a massive hot atmosphere.}; \citealp{Peng2010}) and ram pressure (\citealp{Gunn1972}), that can influence the star formation rate of our central galaxy and that of its accreted satellite galaxies. Ram pressure can both trigger bursts of SF by compressing the gas within a satellite galaxy and quench its SF by stripping it (e.g.,  \citealp{Kapferer2009,Bekki2014, Koutsouridou2019}). These effects might be important for the evolution of Local Group satellites at low $z$ (\citealp{Simpson2018, Hausammann2019} but see also \citealp{salvadori15} for a different view). However, at $z>5$, the main progenitor (or major branch) of the MW with $M_{\rm vir}<10^{11}\:{\rm M_\odot}$ (and $M_*\lesssim 10^9\:{\rm M_\odot}$), is unlikely to hold a sufficiently massive hot atmosphere to exert high ram pressure (see, e.g., \citealp{Gelli2020}). The same is true for the AGN and halo quenching mechanisms, which come about at $M_{\rm vir}\sim10^{12}\:{\rm M_\odot}$, 
or $M_*>10^{10}\:{\rm M_\odot}$  (e.g., ~\citealp{Cattaneo2020, Bluck2020, Koutsouridou2022}). Mergers and disc instabilities can drive inward gas flows that provoke bursts of SF in the central galactic regions \citep{Barnes1991,Teyssier2010, Zolotov2015, Bournaud2016}.
Although the latter can be important for the evolution of high redshift gas-rich galaxies they can be reasonably neglected for a study focused on the stellar halo such as ours.

\subsubsection{The Initial Mass Function of Pop~III and Pop~II/I stars}
\label{sec:stellev}

Following the critical metallicity scenario \citep[e.g.,][]{omukai2000, bromm2001fragmentation, schneider2002first}, we form PopIII stars when the total metallicity, $Z_{\rm gas}$, of the gas within a progenitor halo is below a critical value $Z_{\rm crit} = 10^{-4.5}\:{\rm Z_\odot}$ \citep{deBen2017}, and PopII/I stars whenever $Z_{\rm gas} \geq Z_{\rm crit}$.

We adopt a \citet{larson1998early} IMF for both Pop~III and Pop~II/I stars: 
\begin{equation}
\label{e:Larson}
\phi(m_\star) = \frac{d N}{d m_\star} \propto m_\star^{-2.35} {\rm exp} \bigg( - \frac{m_{\rm ch}}{m_\star} \bigg),
\end{equation}
but with very different characteristic mass, $m_{\rm ch}$, and $m_\star$ range.

For Pop~II stars we assume $m_\star=[0.1,100]\,{\rm M_\odot}$ and $m_{\rm ch}=0.35\:{\rm M_\odot}$, which is consistent with observations of present-day forming stars \citep{Krumholz2015}.

For Pop~III stars we consider $m_\star=[0.1,1000]\,{\rm M_\odot}$ and an IMF biased towards more massive stars, i.e., with $m_{\rm ch}\geq 1 M_\odot$. This mass range and characteristic mass are indeed consistent with constraints on the Pop~III IMF obtained from ultra faint dwarf galaxies \citep{rossi2021ultra} and in line with the results of cosmological hydrodynamical simulations for the formation of Pop~III stars \citep[][see also the Introduction]{Susa2014,hirano2014one,Hirano2015}. In \citet{Pagnini2023}, we showed that a $m_{\rm ch} = 10\:{\rm M_\odot}$ is in good agreement with the observed [C/Fe] range within the bulge and can explain the dearth of CEMP-no stars with [C/Fe]$>+1$ in this environment. Therefore, we adopt $m_{\rm ch}=10\:$M$_{\odot}$ as our starting point and explore the effect of different characteristic masses and a different maximum mass, $m^{\rm max}_{\star}$, for Pop~III stars in Section~\ref{IMF}. 

In some cases, and most commonly in poorly star-forming low-mass halos, the SF bursts are not strong enough to fully populate the theoretical IMF. This has important consequences for the type of stars formed, stellar feedback, chemical evolution and total stellar mass \citep{Kroupa2003,Weidner2006,Weidner2013,deBen2017, Applebaum2020,rossi2021ultra}. To account for that, we implement in {\sc NEFERTITI} a Monte Carlo procedure that generates a random sequence of stars, according to the assumed IMF, with total mass equal to the stellar mass formed in each SF burst (see \citealp{rossi2021ultra} for details). In other words, we account for the incomplete sampling of the IMF of both Pop~III and Pop~II/I stars in poorly star-forming halos.

\subsubsection{The energy distribution of Pop~III SNe}

\begin{figure}
\begin{center}
\includegraphics[width=1\hsize]{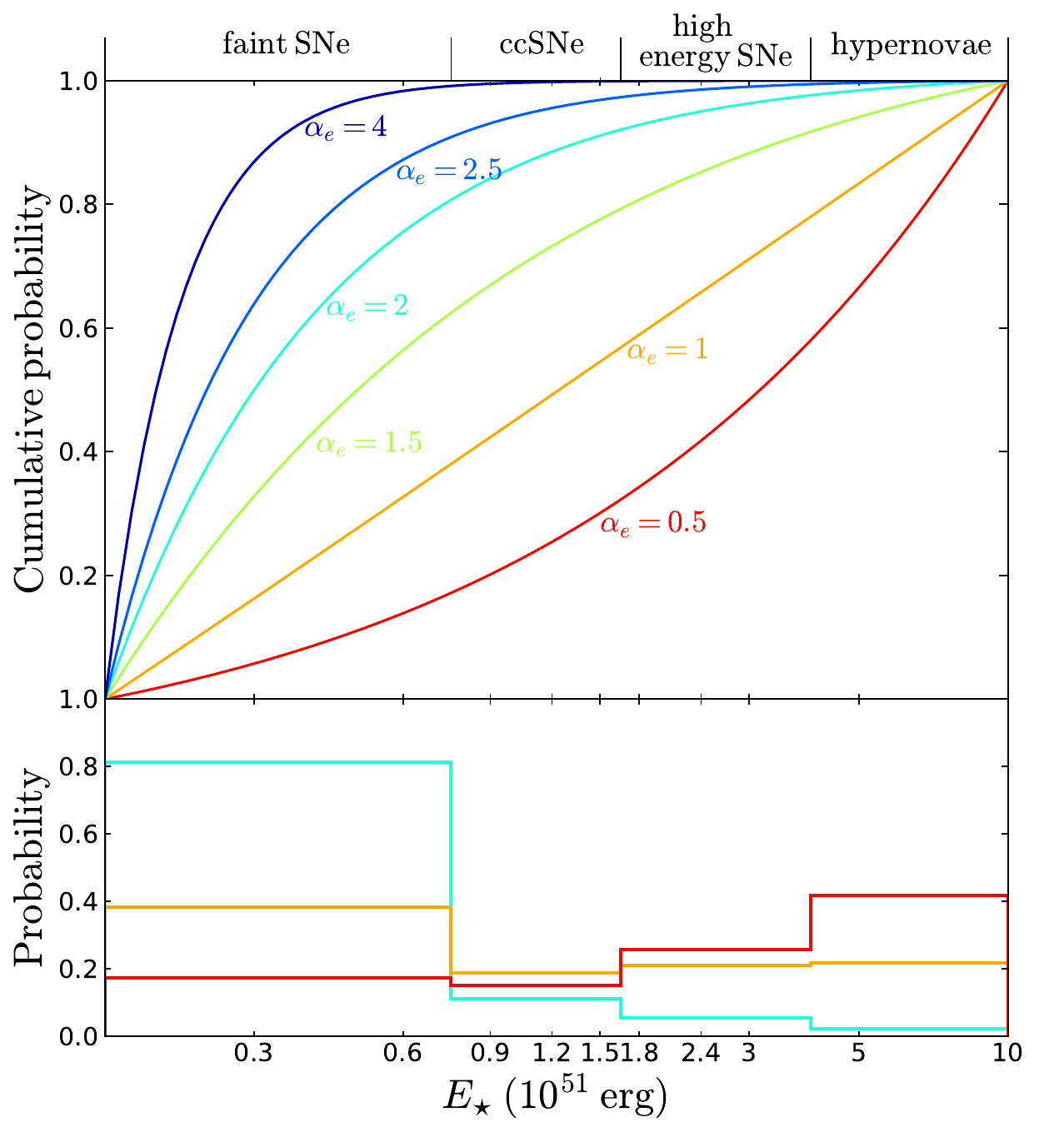} 
\end{center}
\caption{Top: cumulative probability for a Pop~III SN with $m_\star = 10-100\:{\rm M_\odot}$ to have a given explosion energy, $E_{51}$, for different values of the energy distribution function (Eq.~\ref{e:EDF}) exponent, $\alpha_e$. Bottom: the corresponding probability of such star to explode as a faint SN, a ccSN, a high energy SN or a hypernova, for $\alpha_e=0.5$ (red), 1 (orange), and 2 (cyan).}
\label{i:EDF}
\end{figure}

Currently, there is no theoretical constraint on the explosion energies of Pop~III SNe with $m_\star = 10-100\:{\rm M_\odot}$, and while observations suggest that they could have spanned almost 2 orders of magnitude, their distribution remains completely unknown (see~Introduction). Here, we account for the first time for the energy distribution of such primordial SNe in the context of a cosmological galaxy formation model. To this end, we assume a mass-independent energy distribution function (EDF) of the form:
\begin{equation}
\frac{{\rm d}N}{{\rm d}E_\star} \propto E_\star^{-\alpha_e},
\label{e:EDF}
\end{equation}
where $E_\star$ is the explosion energy and $\alpha_e$ is a free parameter of the model. Based on this underlying distribution, we assign randomly an energy level to each Pop~III SN with $m_\star=10-100\:{\rm M_\odot}$. The top panel of Fig.~\ref{i:EDF} shows the cumulative probability as a function of $E_\star$ for the different $\alpha_e$ values considered hereafter. The bottom panel shows the corresponding probability for a Pop~III star to explode as a faint SN ($E_{51} = [0.3, 0.6]$), a core-collapse SN (ccSN; $E_{51} = [0.9, 1.2, 1.5]$), a high energy SN ($E_{51} = [1.8, 2.4, 3]$) or a hypernova ($E_{51} =[5, 10]$), for $\alpha_e=0.5$, 1 and 2.

\label{sec:EDF}

\subsubsection{Relaxing the Instantaneous Recycling Approximation}
In all previous works where the predecessor of our SAM (the {\sc GAMETE} SAM) was coupled with time consuming N-body simulations (\citealp{Salvadori2010, graziani2015galaxy,pacucci2017gravitational,graziani17,Pagnini2023}), the chemical evolution was computed assuming the \textit{instantaneous recycling approximation} (IRA), i.e., that all stars that do not survive until $z=0$, die and return gas and metals into the ISM {\it instantaneously}.  
This approximation overestimates the ISM enrichment rate (steepening the metal-poor tail of the MDF) and describes poorly the abundance of elements produced on long timescales, such as Fe and C. In addition, it blurs out the chemical signatures of different stellar types, such as primordial faint SNe and PISNe, that in reality explode at different times, by mingling their ejecta instantaneously. 

In {\sc NEFERTITI}, we abandon the IRA and, instead, follow the evolution of each individual star, depending on its initial mass and metallicity. At each timestep and for each halo, we compute the rate, $\dot{R}$, at which gas is restored into the ISM through stellar winds or SN explosions from: 
\begin{equation}
\dot{R} = \int_{m_{\rm turn}(t)}^{m^{\rm max}_{\star}} (m_\star - w_m(m_\star)) N(\overline{t}_{\rm form}, m_\star) {\rm d}m_\star,
\label{e:dR}
\end{equation}
 where $N(\overline{t}_{\rm form}, m_\star)$ is the number of stars with mass $m_\star$ that were formed at time $\overline{t}_{\rm form}= t - \tau_\star$, $w_m$ and $\tau_\star$ are the remnant mass and lifetime of a star with initial mass $m_\star$, and $m_{\rm turn}(t)$ is the turnoff mass, i.e., the mass corresponding to $\tau_\star=t$.

Similarly, we define the total ejection rate, $\dot{Y_i}$, of an element~$i$ that is returned to the ISM without being re-processed (first term in the square brackets) and newly produced (second term):
\begin{equation}
\nonumber
\dot{Y_i} = \int_{m_{\rm turn}(t)}^{m^{\rm max}_\star} \big[(m_\star-w_m(m_\star) - m_i(m_\star,Z_\star))Z_i(\overline{t}_{\rm form})+
\end{equation}
\begin{equation}
+m_i(m_\star,Z_\star)\big]N({\overline{t}_{\rm form}}, m_\star){\rm d}m_\star,
\label{e:dY}
\end{equation}
where $m_i(m_\star,Z_\star)$ is the mass of element~$i$ that is synthesized by a star with initial mass $m_\star$ and metallicity $Z_\star$, and $Z_i(\overline{t}_{\rm form})$ is the mass fraction of the element~$i$ at the time of formation of each star. 

We adopt the stellar lifetimes of \citet{raiteri1996simulations} for Pop~II/I stars and the stellar lifetimes of \citet{Schaerer2002} for Pop~III stars. 

\subsubsection{Stellar yields and mixing}
\label{sec:yields_mixing}

The metal yields and remnant masses of Pop~III stars, entering equations~\ref{e:dR} and \ref{e:dY}, are adopted from \citet{Heger2002} for PISNe ($140 \leq m_\star/{\rm M_\odot} \leq 260$) and from \citet{Heger2010} for less massive Pop~III SNe ($10 \leq m_\star/{\rm M_\odot} \leq 100$). The latter are given for 10 different explosion energies in the range $0.3-10\times10^{51}\:{\rm erg}$ (see Section~\ref{sec:EDF}). \citet{Heger2010} use a 1D code that cannot capture mixing between stellar layers due to its multidimensional nature. They, therefore, implement mixing artificially by moving a running boxcar through the star typically four times \citep{Welsh2021}. For each stellar mass and explosion energy, there are 14 different values for the {\it mixing parameter} (in the range $f_{\rm mix}=0-0.2512$), which is defined as the width of the boxcar in units of the helium core mass\footnote{Rotation and mass loss at all stages of stellar evolution are ignored in the \citet{Heger2010} models.}. The mixing parameter is unknown but it can largely affect the abundance of various chemical elements produced by Pop~III SNe, such as carbon, e.g., Vanni et al. in prep. In the absence of theoretical yields for the intermediate pulsational PISNe, i.e., stars with $m_\star = 100-140\:{\rm M_\odot}$, we assume that they collapse into black holes returning no mass into the ISM. 

For Pop~II/I stars we adopt the yields of \citeauthor{Limongi2018} (\citeyear{Limongi2018}; set R without rotation velocity) for massive stars  evolving as core-collapse SNe (ccSNe) and the \citeauthor{vandenHoek1997} (\citeyear{vandenHoek1997}) yields for low and intermediate mass ($m_\star < 8\:{\rm M_\odot}$) Asymptotic Giant Branch (AGB) stars.

\subsubsection{Mechanical feedback from SNe}
\label{sec:mech_fdbk}

Supernovae drive winds that, if sufficiently energetic, can escape the gravitational potential well of their host halo and expel gas and metals into the surrounding medium. Different kinds of SNe are characterised by different explosion energies. For Pop~III PISNe, we adopt the mass-energy relation of \citet{Heger2002}, while for metal-free stars with $m_\star$ = 10-100${\:{\rm M_\odot}}$ we consider all energy levels ($E_{51}=0.3 - 10$) provided by \citet{Heger2010}, independently of the stellar mass as explained in Section~\ref{sec:EDF}. For Pop~II/I ccSNe, we assume an average explosion energy of $10^{51}\:$erg. Therefore, at each timestep, the total power output from SNe in a halo is $\sum\limits_{i} \dot{N}_{\rm SN}^i \cdot E_{\rm SN}^i$, where $\dot{N}_{\rm SN}^i$ is the explosion rate of SNe with energy $E_{\rm SN}^i$. If a fraction $\epsilon_{\rm wind}$ of this power is converted into kinetic form, the gas outflow rate from the halo, $\dot{M}_{\rm gas,ej}$, will satisfy:
\begin{equation}
\frac{1}{2} \dot{M}_{\rm gas,ej} u_{\rm esc}^2 = \epsilon_{\rm wind} \sum\limits_{i} \dot{N}_{\rm SN}^i E_{\rm SN}^i, 
\label{e:Mej}
\end{equation}
where the wind efficiency, $\epsilon_{\rm wind}$, is the second free parameter of our model and $u_{\rm esc}=\sqrt\frac{GM_{\rm vir}}{r_{\rm vir}}=f(M_{\rm vir}, z)$ \citep{barkana2001beginning} is the escape speed of the halo.

\subsubsection{Following the gas evolution}

At each sub-timestep $\delta t_{\rm s}$ of the SAM, we compute the evolution of the total gas mass, $M_{\rm gas}$ and mass of element~$i$, $M_{Z_i}$ in the ISM of each galaxy from the equations:
\begin{equation}
\dot{M}_{\rm gas} = \dot{M}_{\rm gas, accr} - {\rm SFR} + \dot{R} - \dot{M}_{\rm gas, ej}
\end{equation}
and 
\begin{equation}
\label{e:Z_ISM}
\dot{M}_{{\rm Z}_i} = Z_i^{\rm IGM}\dot{M}_{\rm gas, accr} - Z_i {\rm SFR} + \dot{Y}_i - Z_i\dot{M}_{\rm gas, ej},
\end{equation}
respectively, where $Z_i = M_{Z_i}/M_{\rm gas}$ is the mass fraction of element~$i$ in the ISM and $Z_i^{\rm IGM}$ is its mass fraction in the IGM. The latter is updated after each sub-timestep by summing the contributions of all haloes~$h$:
\begin{equation}
\label{e:Z_IGM}
{\dot{M}_{Z_i}}^{\rm IGM} = \mathlarger{\mathlarger{\sum}}\limits_{h} \Big(-Z_i^{\rm IGM}\dot{M}_{\rm gas, accr}^h +Z_i^h\dot{M}_{\rm gas, ej}^h \Big).
\end{equation}
Equations~\ref{e:Z_ISM} and \ref{e:Z_IGM} imply that the ejecta of dying stars are instantaneously and homogeneously mixed within both the ISM of each halo and the IGM, which are thus characterized by a unique chemical composition at each timestep. This perfect mixing approximation has significant consequences for the chemical evolution of our system, which are discussed in Sections~\ref{Results} and \ref{Discussion}.

\subsection{Model calibration}
\label{calibration}

\begin{figure}
\begin{center}
\includegraphics[width=1\hsize]{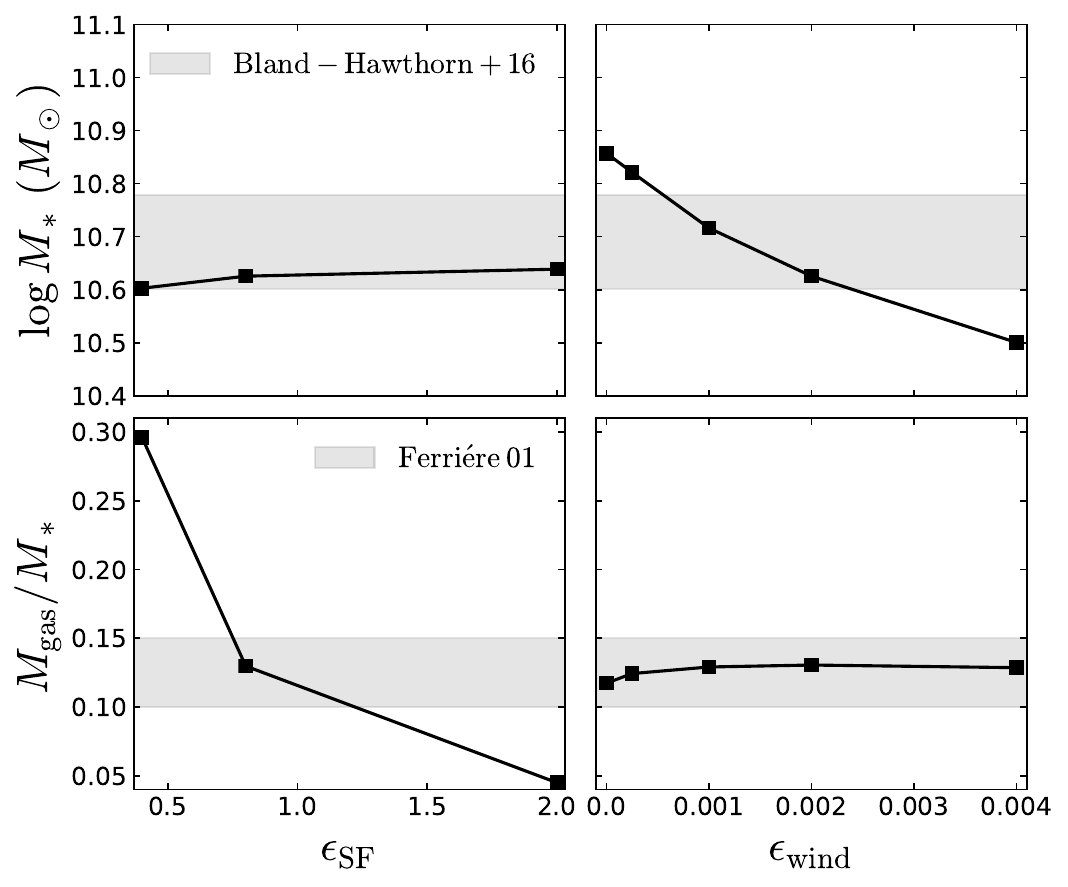} 
\end{center}
\caption{Total stellar mass (top row) and gas-to-stellar mass ratio (bottom row) of our MW-analogue at $z=0$, as a function of the two free parameters of our model: the star formation efficiency (for $\epsilon_{\rm wind}=0.002$; left) and the wind efficiency (for $\epsilon_{\rm SF}=0.8$; right). Gray shaded areas represent the observed global properties of the MW (see text for details).}
\label{i:cali}
\end{figure}

In Fig.~\ref{i:cali}, we compare the observed global properties of the MW (grey shaded areas) with the results of our model at $z=0$, as a function of the SF efficiency $\epsilon_{\rm SF}$ (left), and the wind efficiency $\epsilon_{\rm wind}$ (right panels). One can see that the final stellar mass, $M_*$, depends strongly on $\epsilon_{\rm wind}$ but only weakly on $\epsilon_{\rm SF}$, while the opposite is true for the gas-to-stellar mass ratio, $M_{\rm gas}/M_*$. This can also be inferred by solving analytically the equations that govern the evolution of the stellar and gas mass in a galaxy (see equations~4 and 6 in \citealp{Koutsouridou2019}). Therefore, these two observables are sufficient to constrain the two free parameters of our model. 

We adopt $\epsilon_{\rm SF}=0.8$, and $\epsilon_{\rm wind}=0.002$, for which we obtain $M_* \approx 4.2 \times 10^{10}\:{\rm M_\odot}$, and $M_{\rm gas}/M_*\approx0.13$ at $z=0$. 
\citet{Bland2016} report $M_* = (5 \pm 1) \times 10^{10}\:{\rm M_\odot}$ (gray areas in the top row of Fig.~\ref{i:cali}), and $M_{\rm vir} =(1.3 \pm 0.3)\times 10^{12}\:{\rm M_\odot}$ for the Milky Way, by combining estimates from dynamical model fitting to stellar surveys and to the Galactic rotation curve (see references therein). Using the same approach, \citet{McMillan2017} and \citet{Cautun2020} find similar values: $M_* = (5.43 \pm 0.57) \times 10^{10}\:{\rm M_\odot}$, $M_{\rm vir} =(1.3 \pm 0.3) \times 10^{12}\:{\rm M_\odot}$; and $M_* = 5.04^{+0.43}_{-0.52} \times 10^{10}\:{\rm M_\odot}$, $M_{\rm vir} =1.08^{+0.20}_{-0.14} \times 10^{12}\: {\rm M_\odot}$, respectively\footnote{Studies estimating the MW stellar mass from direct integration of starlight usually find higher values, $M_*\sim6 \times 10^{10}\:{\rm M_\odot}$ \citep{Bland2016}, but these studies do not simultaneously constrain its virial mass.}. We choose a value for $\epsilon_{\rm wind}$ that gives a stellar mass at the lower limit of the observational range, since the virial mass of our system is lower than most observational estimates (see Section~\ref{DM}). By substituting $\epsilon_{\rm wind}=0.002$ in Eq.~\ref{e:Mej} we get a mass loading factor\footnote{Note that $u_{\rm esc}^2 \approx 2.8 \times 10^{47}\: {\rm erg/M_\odot}$ for $M_{\rm vir}=7.8 \times 10^{11}\:{\rm M_\odot}$ at $z=0$ (Eq.~27 in \citealp{barkana2001beginning})  and that with the assumed Pop~II/I IMF we form one SN every $100\:{\rm M_\odot}$.} $\eta \equiv \dot{M}_{\rm gas,ej}/{\rm SFR} \approx 0.14$ for our MW-analogue at $z=0$, in agreement with recent observational estimates ($\eta_{\rm MW}=0.1 \pm0.06$, \citealp{Fox2019}).

 Our adopted value for $\epsilon_{\rm SF}$ results in a gas-to-stellar mass ratio within the typically reported range of 0.1-0.15 \citep{Ferriere2001,Stahler2004}. In addition, at $z=0$, $\epsilon_{\rm SF}$ corresponds to a star formation timescale $t_{\rm SF} \equiv M_{\rm gas}/{\rm SFR} \sim1.9\:{\rm Gyr}$, and a SFR$\sim2.7\:{\rm M_\odot/yr}$ for our MW-analogue, in agreement with observational estimates ($t_{\rm SF, MW} = 2\:{\rm Gyr}$, \citealp{Bigiel2008}; ${\rm SFR_{MW}} = 1-3\:{\rm M_\odot/yr}$, \citealp{Chomiuk2011,Bland2016}). At $z=0$, the mean metallicity of the ISM, $Z_{\rm ISM} = 1.07\:Z_\odot$, in our MW-analogue and of the IGM, $Z_{\rm IGM}\sim0.17\:Z_\odot$, are in accordance with observations in the Galactic disc and in high-velocity clouds currently accreting onto the disc ($\sim0.1-0.3\:{\rm Z_\odot}$; \citealp{Ganguly2005,Tripp2003,Danforth2008}). For all the above reasons, we are confident that our model with the selected values for $\epsilon_{\rm SF}$ and $\epsilon_{\rm wind}$, is a good representation of the evolution of a MW-like galaxy and its immediate environment.

\section{Stellar data for model comparison}
\label{Data}

This Section describes in detail the available observations of Galactic halo stars that we use to compare with key results of our model -- namely the predicted metallicity distribution function, the fraction of CEMP stars, the carbonicity distribution function and the distribution of stars in the [C/Fe]--[Fe/H] diagram.

\begin{itemize}   
\item {\it The Metallicity Distribution Function (MDF):} for Galactic halo stars with $-4<\rm[Fe/H]<-2$ we adopt the MDF proposed by \citet{Bonifacio2021}, which is the largest and most complete (i.e., biased-corrected) MDF for model comparison. It represents the average of three independently derived MDFs, the one by \citeauthor{Naidu2020} (\citeyear{Naidu2020}; H3 Survey), the uncorrected one by \citeauthor{Schorck2009} (\citeyear{Schorck2009}; Hamburg/ESO Survey) and the one determined by \citet{Bonifacio2021} themselves, and corrected for selection biases, from Sloan Digital Sky Survey spectra. The standard deviation in each metallicity bin (shown with the black errorbars in Fig.~\ref{i:mdf}) is provided by \citet{Bonifacio2021} as an error estimate on the MDF. The three MDFs used for computing the average are essentially identical above $\rm[Fe/H]\sim-3$, explaining the small errors in this metallicity range. However, there are other published Galactic halo MDFs, that appear steeper (e.g., the corrected MDF by \citealp{Schorck2009}, and the one of \citealp{Carollo2010}) or shallower (e.g.,~\citealp{youakim2020pristine}) in this [Fe/H] range. At lower [Fe/H], the \citet{Naidu2020} MDF, which is based on high-resolution data, is undefined. The \citet{Schorck2009} MDF and the one determined by \citet{Bonifacio2021} extend down to $\rm[Fe/H]\sim-4$, but are based on low resolution ($R\sim2000$) data. Due to the fact that metallicities of $\rm[Fe/H]<-3$ can only be accurately and precisely determined through high-resolution spectra, and due to the low number statistics at low [Fe/H], the average MDF shows much larger errors at $-4\leq {\rm [Fe/H]}\leq-3$.  
\\
\item {\it The lowest-Fe tail of the MDF:}  we compute the halo MDF at $\rm [Fe/H]<-4$ using data from the SAGA\footnote{\url{http://sagadatabase.jp/}} database (2023, April 10 version), which assembles the abundances of all $\rm[Fe/H]\leq -2.5$ stars derived from high and medium-resolution follow up observations \citep{Suda2008, Suda2011, Yamada2013}. As researchers usually favour detailed studies of the most extreme stars, follow-up observations are biased towards low metallicities. For this reason, we only consider the SAGA MDF at $\rm[Fe/H]<-4$ (including 43 stars), where we can assume that follow-up is near-complete. Still, it is likely that not all known stars with estimated $\rm-4.5\leq[Fe/H]\leq-4$ have been followed-up at high resolution. If those are confirmed in the future, the number of stars at ${\rm [Fe/H]}<-4.5$ with respect to the number of stars at $-4.5 \leq {\rm [Fe/H]} \leq -4$ will decline. We represent this possibility qualitatively with down-pointing arrows (shown, for example, in Fig.~\ref{i:mdf2}).
\\
\item {\it The fraction of CEMP-no stars:}  we compute the fraction of CEMP-no stars,
\begin{equation}
\label{e:F_CEMP}
F_{\rm CEMP}({\rm [Fe/H]}) = \frac{N_{\rm CEMP}({\rm [Fe/H]})}{N_*([{\rm Fe/H}])},
\end{equation}

where $N_{\rm CEMP}$ is the number of CEMP-no stars, and $N_*$ the total at a given [Fe/H] bin,  
using the high/medium-resolution sample of \citeauthor{placco2014carbon} (\citeyear{placco2014carbon}) and the high-resolution sample of \citeauthor{Yong2013a} (\citeyear{Yong2013a}). \citet{placco2014carbon} collected a large sample of VMP literature stars, excluded those with $\rm[Ba/Fe]>+0.6$ and $\rm[Ba/Sr]>0$, which were likely enriched by an AGB companion (CEMP-s stars), and corrected the carbon abundances of the remaining sample to account for evolutionary effects. Their estimated fractions of CEMP-no ($\rm[C/Fe]\geq+0.7$) stars are shown in Fig.~\ref{i:cemp}.  
\citet{Yong2013a} performed a homogeneous chemical abundance analysis of 190 literature and program stars, of which 172 have $\rm[Fe/H]<-2$. We completed their catalogue at low metallicities by adding the more recently discovered EMP stars shown in Fig.~\ref{i:ssp} (diamond points) and listed below. 
We computed $N_{\rm CEMP}$ using the \citet{Yong2013b} classification of CEMP-no stars (that is based on the \citealp{Aoki2007} criterion\footnote{Both the \citet{Aoki2007} criterion for CEMP stars ($\rm[C/Fe]\geq+0.70$, for ${\rm log}(L/L_\odot)\leq2.3$ and $\rm[C/Fe]\geq+3.0 - {\rm log}(L/L_\odot)$, for ${\rm log}(L/L_\odot)>2.3$) and the corrections  of \citet{placco2014carbon}, account for the depletion of the surface carbon abundance of stars as they ascent the red-giant branch.}) and excluding stars with $\rm [Ba/Fe]>+0.6$. The latter criterion was adopted to be consistent with \citet{placco2014carbon}. In both observational surveys the CEMP-no fraction decreases with increasing [Fe/H]. However, at $\rm-4<[Fe/H]<-3$, the $F_{\rm CEMP}$ of \citet{Yong2013a} are significantly lower than those of \citeauthor{placco2014carbon} (\citeyear{placco2014carbon}; see Fig.~\ref{i:cemp}) highlighting the uncertainties in the observational estimates of $F_{\rm CEMP}$ (see also Section~\ref{Conclusion}).
\\
\item {\it The Carbonicity Distribution Function (CDF):} we compare our predictions for the [C/Fe] ("carbonicity", \citealp{Carollo2012,Lee2017}) distribution function of VMP inner halo stars, to observations from the SAGA database. We do so only at $\rm[C/Fe]>+2$ where we expect the observational sample to have higher completeness. We compute the SAGA CDF by including all confirmed CEMP-no stars, CEMP stars with upper limits for barium enhancement at $\rm[Ba/Fe]>+0.6$ as well as CEMP stars with no measurement of barium abundances (35 stars in total with $\rm [C/Fe]>+2$ and $\rm [Fe/H]\leq-2$, including upper limits for [C/Fe]). 
\\
\item {\it The [C/Fe] vs [Fe/H] abundances:}  in the left panel of Fig.~\ref{i:ssp} we compare our predicted distribution of halo stars in the [C/Fe]--[Fe/H] diagram with the CEMP-no and C-normal stars in the catalogues of \citeauthor{Yong2013a} (\citeyear{Yong2013a}; X points) and \citeauthor{placco2014carbon} (\citeyear{placco2014carbon}; circles) as well as with individual stars (diamond points) by \citet{Christlieb2002, Norris2007, Caffau2011, Keller2014,Hansen2014, Frebel2015,Li2015, Bonifacio2015, Caffau2016, Bonifacio2018, Francois2018,  Aguado2018b, Starkenburg2018, Aguado2019, Ezzeddine2019} and \citeauthor{Nordlander2019} (\citeyear{Nordlander2019})\footnote{The \citet{Norris2007} star and the \citet{Aguado2019}/\citet{Ezzeddine2019} star are already included in the sample of \citet{Yong2013a} and were, therefore, not added to it before computing $F_{\rm CEMP}$.}. 

\end{itemize}

\section{Results}
\label{Results}
Using our model (Section~\ref{sec:model}), we investigate how the MDF, the CDF and the fraction of CEMP-no stars in the Galactic halo depends on the unknown energy distribution function (EDF) of the first Pop~III supernovae. In Section \ref{IMF} we examine the degeneracy between the EDF and the IMF of Pop~III stars and in Section~\ref{POPIII} we determine the key observables to constrain them.

Our findings are impacted by the stochastic sampling of both the masses and the SN explosion energies of Pop~III stars (Section~\ref{sec:stellev}). Therefore, for each model presented in this article, we have averaged over 100 realizations as we find that this number is sufficient for our results to converge. We focus on the surviving VMP ($\rm[Fe/H]\leq-2$) stars that lie in the inner Galactic halo, i.e. at galactocentric radii $7\:{\rm kpc} \leq R_{\rm gal} \leq 20\:{\rm kpc}$, at $z=0$. Due to the instantaneous mixing of the IGM (that reaches $\rm[Fe/H]\approx -2$, by $z=5$), we find that all VMP stars are formed before $z\sim5$, regardless of the assumed properties (EDF, IMF and mixing) of Pop~III stars. Since we do not consider binary mass transfer, all CEMP (and C-normal) stars in our models, reflect the abundances of their birth clouds. We adopt the solar abundances from \citet{Asplund2009}.

\subsection{Pop~III stars: Energy Distribution Function vs IMF}
\label{IMF}

This Section explores how the properties of the surviving inner halo stars in our model change when we vary the EDF of the first SNe. In what follows, we assume a flat distribution for the unknown Pop~III mixing parameter and assign randomly one of the 14 mixing levels provided by \citet{Heger2010} to each Pop~III star with $10\leq m_\star/{\rm M_\odot}\leq 100$. 
We consider four Larson-type IMFs (Eq.~\ref{e:Larson}): (i) with characteristic mass $m_{\rm ch}=1\:{\rm M_\odot}$ and maximum mass for Pop~III stars $m_\star^{\rm max}=1000\:{\rm M_\odot}$, (ii) with $m_{\rm ch}=10\:{M_\odot}$ and $m_\star^{\rm max}=1000\:{\rm M_\odot}$, (iii) with $m_{\rm ch}=100\:{M_\odot}$ and $m_\star^{\rm max}=1000\:{\rm M_\odot}$ and (iv) with $m_{\rm ch}=10\:{M_\odot}$ and $m_\star^{\rm max}=100\:{\rm M_\odot}$.

\begin{figure*}
\begin{center}
\includegraphics[width=0.76\hsize]{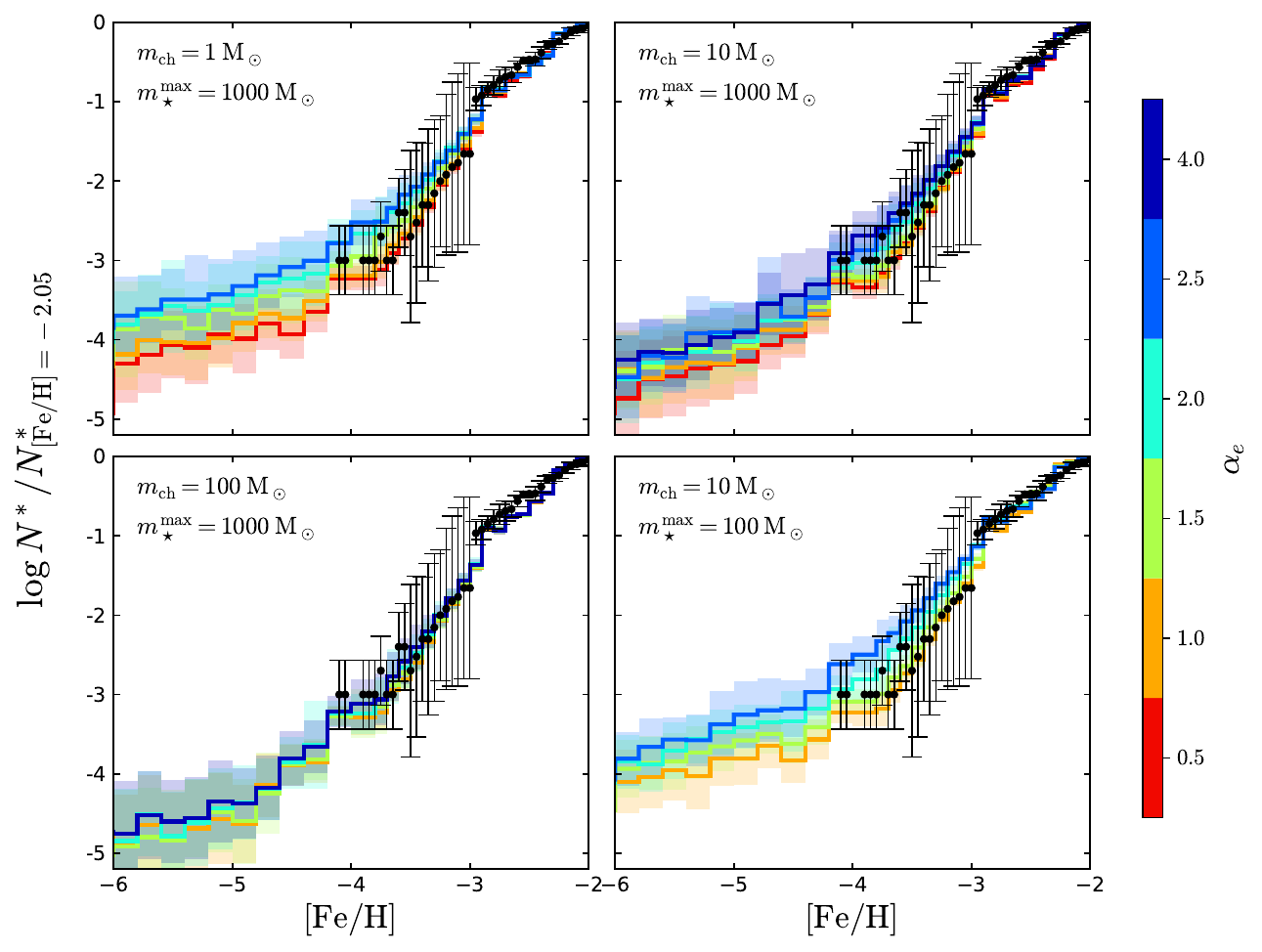}
\end{center}
\caption{Mean metallicity distribution functions of VMP inner halo stars, normalized at $\rm[Fe/H]=-2.05$, in comparison to the observations by \citeauthor{Bonifacio2021} (\citeyear{Bonifacio2021}; points with errorbars). Four Larson-type (Eq.~\ref{e:Larson}) IMFs for Pop~III stars are considered: the first three with mass range $m_\star = 0.1-1000\:{\rm M_\odot}$ and characteristic mass $m_{\rm ch}=1\:{\rm M_\odot}$ (top-left), $m_{\rm ch}=10\:{\rm M_\odot}$ (top-right) and $m_{\rm ch}=100\:{\rm M_\odot}$ (bottom-left) and the fourth with $m_\star = 0.1-100\:{\rm M_\odot}$ and $m_{\rm ch}=10\:{\rm M_\odot}$ (bottom-right panel). For each IMF, results are shown for different values of the Pop~III EDF exponent (see Eq.~\ref{e:EDF}), $\alpha_e$, as denoted by the color. All mixing values of the \citealp{Heger2010} yields are assumed to be equally probable.}
\label{i:mdf}
\end{figure*}

\begin{figure*}
\begin{center}
\includegraphics[width=0.7\hsize]{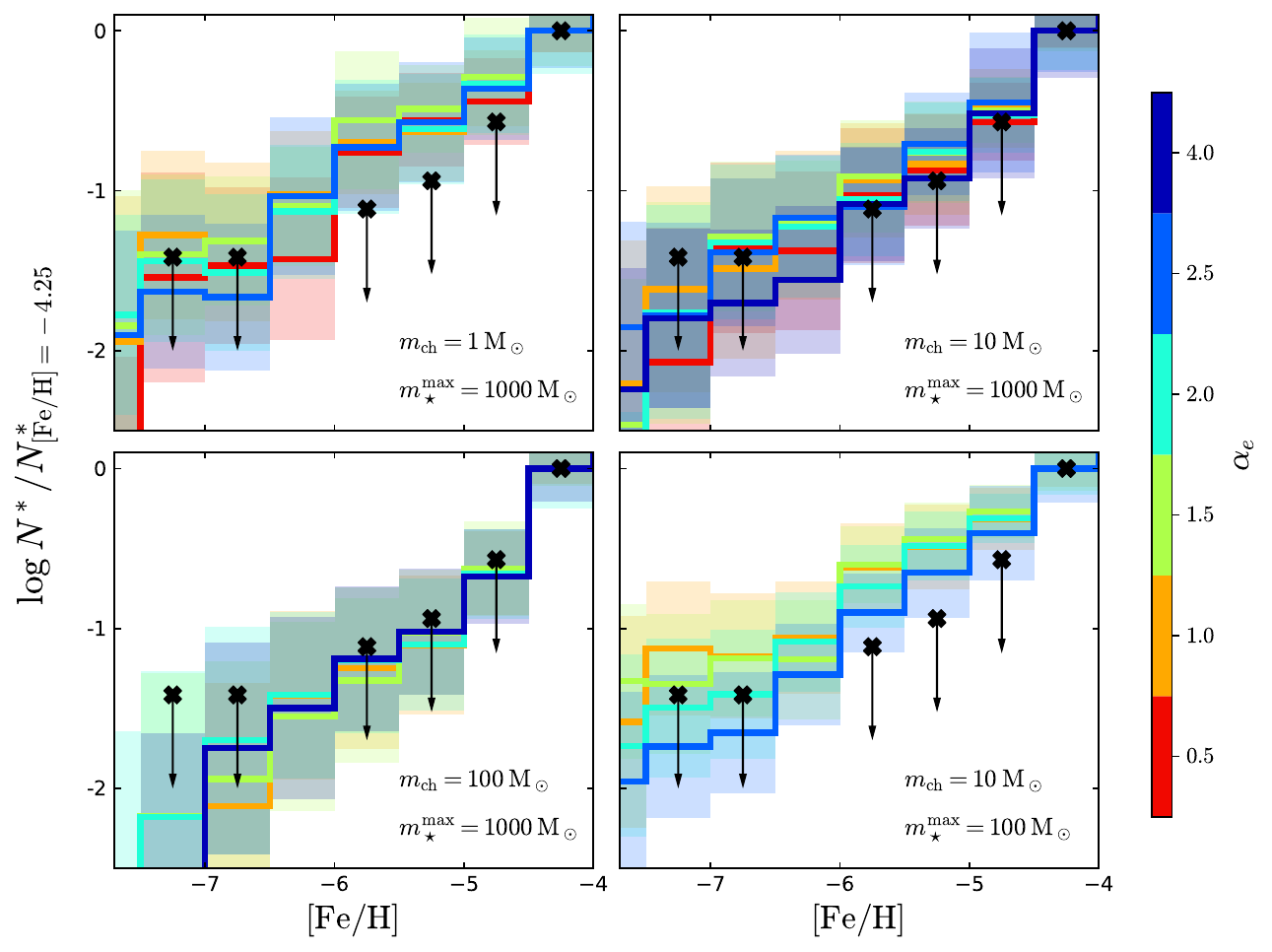}
\end{center}
\caption{Same as Fig.~\ref{i:mdf} but normalized to $\rm[Fe/H]=-4.25$ and compared to the ultra metal poor MDF from the SAGA database. The downwards pointing arrows indicate the possibility that the SAGA MDF might be biased towards the lowest metallicities (see Section~\ref{Data}).}
\label{i:mdf2}
\end{figure*}

\begin{figure*}
\begin{center}
\includegraphics[width=0.76\hsize]{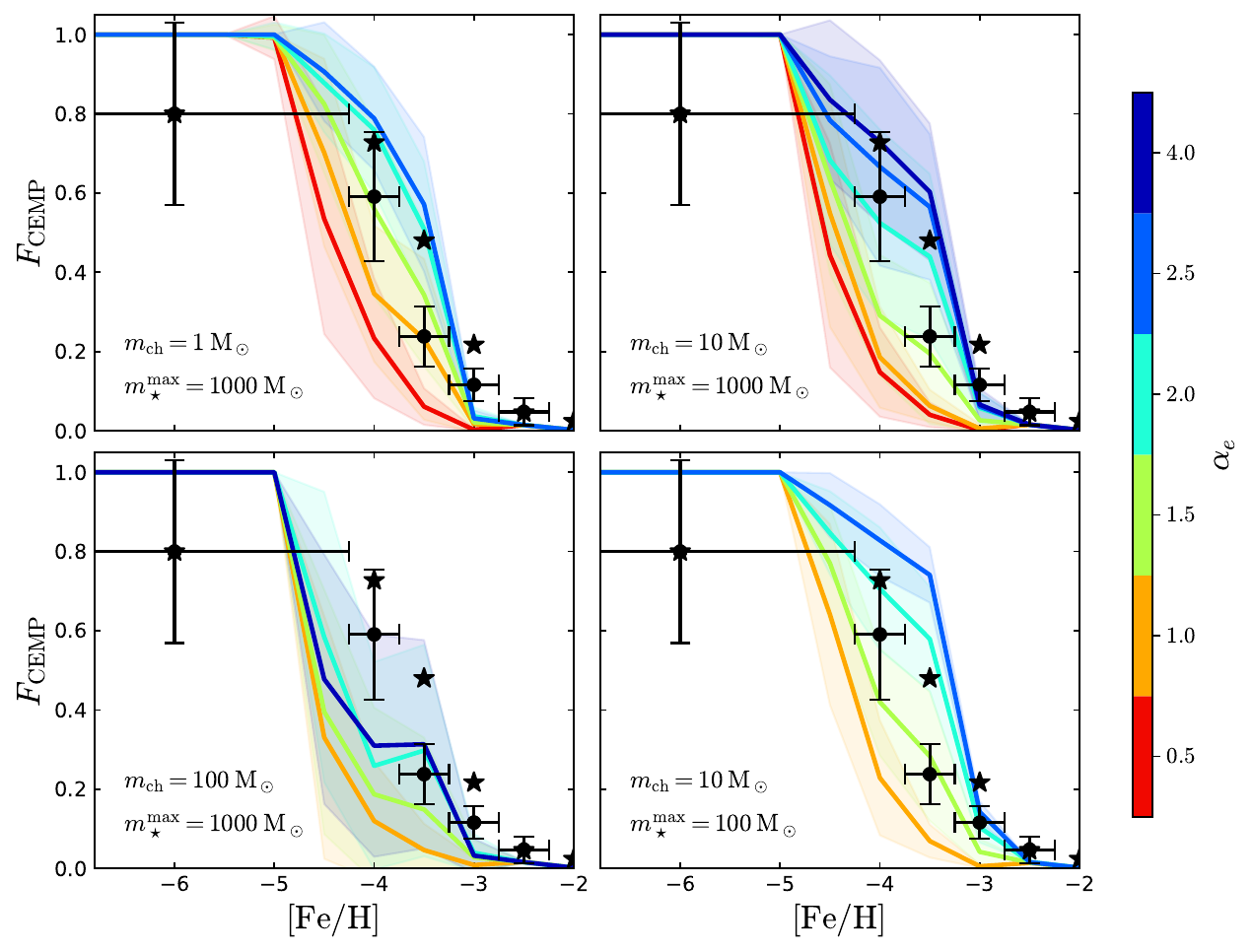}
\end{center}
\caption{Differential CEMP fractions of very metal poor inner halo stars, for the four IMFs and different $\alpha_e$ values considered (see Section~\ref{IMF} or caption of Fig.~\ref{i:mdf}). Datapoints show the observational estimates of \citeauthor{Yong2013a} (\citeyear{Yong2013a}; points with errorbars) and \citeauthor{placco2014carbon} (\citeyear{placco2014carbon}; stars).}
\label{i:cemp}
\end{figure*}

\begin{figure*}
\begin{center}
\includegraphics[width=0.76\hsize]{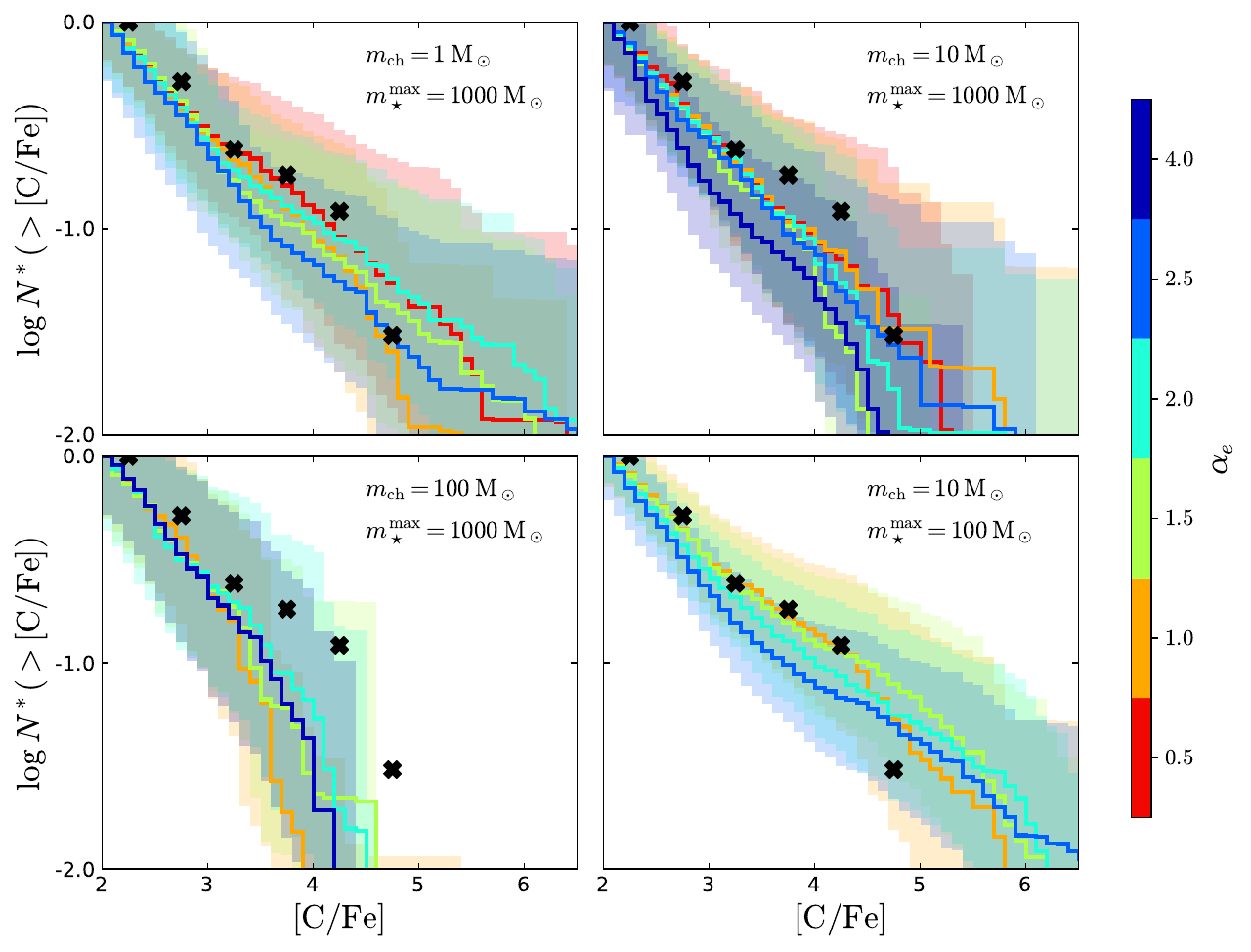}
\end{center}
\caption{Cumulative [C/Fe] distribution functions of very metal poor, inner halo stars, normalized at [C/Fe]=+2, in comparison with observations from the SAGA database.}
\label{i:cdf2}
\end{figure*}

Fig.~\ref{i:mdf} shows the predicted MDFs for the four IMFs assuming different values for the EDF (Eq.~\ref{e:EDF}) exponent as denoted by the color (see Fig.~\ref{i:EDF}). We find that at $\rm[Fe/H]>-3$, the halo MDF is essentially independent of the assumed Pop~III IMF and EDF and is in almost perfect agreement with the one derived by \citet{Bonifacio2021}. It is worth noting that the MDF is a genuine prediction of our model since no free parameter was tuned to reproduce it (Section~\ref{calibration}). At $\rm[Fe/H]<-3$, the MDFs resulting from different models start to differentiate and a clear trend emerges: the MDF steepens as we move both to a higher characteristic mass $m_{\rm ch}$ (or higher $m_\star^{\rm max}$) at fixed EDF or to a lower $\alpha_e$ at fixed IMF, i.e., as more Pop~III stars with high masses and explosion energies are formed. This is to be expected, since high energy SNe, hypernovae and massive PISNe yield more iron than faint and ccSNe and, thus, accelerate the chemical enrichment in their host halos resulting in a steeper MDF. Nevertheless, all MDFs are in agreement with the \citet{Bonifacio2021} one within the (large) errorbars. 

In addition, all model MDFs are in agreement with the SAGA MDF at $\rm[Fe/H]<-4$ within errors (Fig.~\ref{i:mdf2}). Yet, the ones predicted by the $[m_{\rm ch}, m_\star^{\rm max}] = [1,1000]\:{\rm M_\odot}$ and the $[m_{\rm ch}, m_\star^{\rm max}] = [10,100]\:{\rm M_\odot}$ models, which result in less (or zero) PISNe formed, lie above the observational datapoints in most metallicity bins. Instead the ones with $m_{\rm ch}=10\:{\rm M_\odot}$ and $m_{\rm ch}=100\:{\rm M_\odot}$ (and $m_\star^{\rm max}=1000\:{\rm M_\odot}$) lie on top and below the observations, respectively. As explained in Section~\ref{Data}, it is probable that the number of stars at $\rm[Fe/H]<-4.5$ with respect to the number of stars at $\rm[Fe/H]\approx-4$ will decline in the future as more stars in the range $-4.5\leq{\rm [Fe/H]}\leq -4$ will be followed up with high resolution observations. Therefore, we could say that the latter two models are preferable here, or else that PISNe are required to steepen the ultra metal poor tail of the MDF. However, this could be a premature conclusion, given that only 14 stars with $\rm[Fe/H]<-4.5$ have been observed to date.

In Fig.~\ref{i:cemp}, we compare our predicted CEMP fractions to the observations of \citet{Yong2013a} and \citeauthor{placco2014carbon} (\citeyear{placco2014carbon}; see Section~\ref{Data}). All halo stars with $\rm[Fe/H]<-5$ are predicted to be carbon enhanced, $\rm[C/Fe]>+0.7$. At higher metallicities and for a given Pop~III IMF, we find that EDFs skewed towards high explosion energies (i.e. with smaller $\alpha_e$, see Fig.~\ref{i:EDF}) result in lower CEMP fractions; naturally, since more energetic Pop~III SNe yield less [C/Fe] at fixed mass and mixing level (e.g. see Vanni et al. in prep). 
 
The dependence of the yielded [C/Fe] on the Pop~III stellar mass is not straightforward. At $10\leq m_\star/{\rm M_\odot} \leq 100$ and at a given explosion energy, the ejected [C/Fe] appears to increase with mass, especially at low mixing levels, but the relation is not monotonous and tends to reverse at the highest explosion energy $E_\star = 10 \times 10^{51}\:{\rm erg}$ \citep{Heger2010}. The opposite is true for PISNe; the yielded [C/Fe] decreases dramatically with stellar mass from $\sim 10^{13}$ at $m_\star = 140\:{\rm M_\odot}$ to $\le 10^{-1}$ at $m_\star = 260\:{\rm M_\odot}$ \citep{Heger2002}. In addition, only the lowest explosion energies ($E_{51}\leq 1.5$) and the lowest mixing levels produce [C/Fe] that exceed those of the least massive PISNe ($m_\star\lesssim 170\:{\rm M_\odot}$). Yet, all non-PISNe yield higher [C/Fe] than PISNe with $m_\star\gtrsim 195\:{M_\odot}$. 

Nevertheless, Fig.~\ref{i:cemp} reveals a clear trend. As we increase the characteristic mass from $m_{\rm ch}=1\:{M_\odot}$, to $m_{\rm ch}=10\:{M_\odot}$, and $m_{\rm ch}=100\:{M_\odot}$  (resulting in $M_{\rm PISN}/M_{\rm PopIII} \approx 0.04, 0.11$, and 0.22, respectively) the predicted CEMP fraction for a given EDF decreases. For $m_{\rm ch}=1\:{\rm M_\odot}$ we can reproduce the observed $F_{\rm CEMP}$ for $\alpha_e>1$ while for $m_{\rm ch}=10\:{\rm M_\odot}$ we need a higher $\alpha_e \gtrsim 1.5$. That means that as the number of PISNe increases, the number of hypernovae should drop (from $<22\%$ for $m_{\rm ch}=1\:{\rm M_\odot}$ to $<8\%$ for $m_{\rm ch}=10\:{\rm M_\odot}$). 
For $m_{\rm ch}=100\:{\rm M_\odot}$ even the EDF with $\alpha_e=4$ (giving $\sim99\%$ faint SNe) cannot produce enough CEMP stars to meet the observations. This is because when $m_{\rm ch}=100\:{\rm M_\odot}$, PISNe dominate the ISM enrichment thus washing out the high [C/Fe] yielded by faint SNe \citep{Pagnini2023}. Indeed, in the case where $m_\star^{\rm max}= 100\:{\rm M_\odot}$, i.e., when no PISNe are allowed to form, our model fits the observations for $\alpha_e\sim1.5-2$. 

 Finally, in Fig.~\ref{i:cdf2}, we compare the cumulative CDFs, for stars with [Fe/H]$\leq-2$, predicted by our models with observations from the SAGA database. We find that the CDFs become steeper as the $m_{\rm ch}$ of the Pop~III IMF increases, or else as more PISNe form. Models with $m_{\rm ch}=100\:{\rm M_\odot}$ (bottom left panel) significantly underpredict the number of stars with [C/Fe]>+4, yet all other models are in agreement with the observations within errors. At fixed IMF, we see that when the model CDFs are normalized to [C/Fe]=+2, they show no clear dependence on the assumed EDF. That is not true, however, when we consider the CDFs extending down to lower [C/Fe]. There, our now-familiar trend is evident; the higher the energy of Pop~III SNe (or else the lower the $\alpha_e$), the lower the yielded [C/Fe] and, therefore, the steeper the resulting CDF (see Appendix~\ref{appendix1}).


\begin{center}
\captionof{table}{Pop~III stellar parameters for the models that successfully reproduce the observed MDF \citep{Bonifacio2021}, the CDF and the CEMP fractions \citep{Yong2013a, placco2014carbon} of very metal-poor stars in the inner halo. In all models below, all values of stellar mixing given by \citet{Heger2010} are assumed to be equally probable.}
\renewcommand{\cellset}{\renewcommand{\arraystretch}{1.8}}
\begin{tabular}{l | l  | l l l}
\addlinespace
\addlinespace
\hline
 & & Model 1 & Model 2 & Model 3 \\
 \hline
IMF & \makecell{$m_{\rm ch}$\\ $m_\star^{\rm max}$}& \makecell{$1\:{\rm M_\odot}$\\ $1000\:{\rm M_\odot}$} & \makecell{$10\:{\rm M_\odot}$\\ $1000\:{\rm M_\odot}$} & \makecell{$10\:{\rm M_\odot}$\\ $100\:{\rm M_\odot}$}  \\
 & & & &\\
EDF & $\alpha_e$ & 1.0-2.0 & 1.5 - 2.5 & 1.5-2.0  \\

\end{tabular}
\vspace{\dimexpr0.9\baselineskip-\parskip}
 
\label{t:t1}
\end{center}

\subsection{Metal contribution from Pop~III stars}
\label{POPIII}

In the previous Section, it became clear that there exist degeneracies between the EDF and the IMF of Pop~III stars. Furthermore, given the currently large errors in the observational data, it is hard to single out a preferred model. Therefore, it is useful to examine the average behaviour of all our "successful" models, see Table~\ref{t:t1}, i.e., those that are in better agreement with the observed MDF, CDF and $F_{\rm CEMP}$.

\begin{figure*}
\begin{center}$
\begin{array}{cc} 
\includegraphics[width=0.48\hsize]{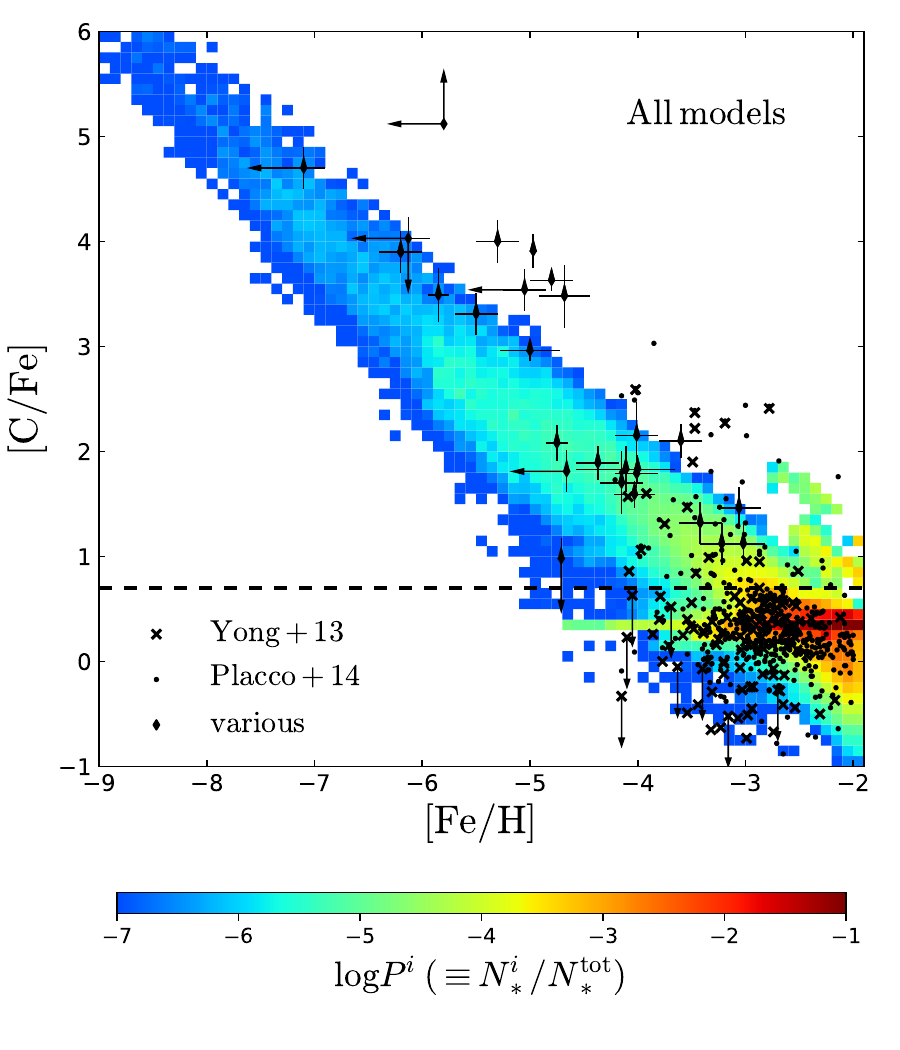} &
\includegraphics[width=0.48\hsize]{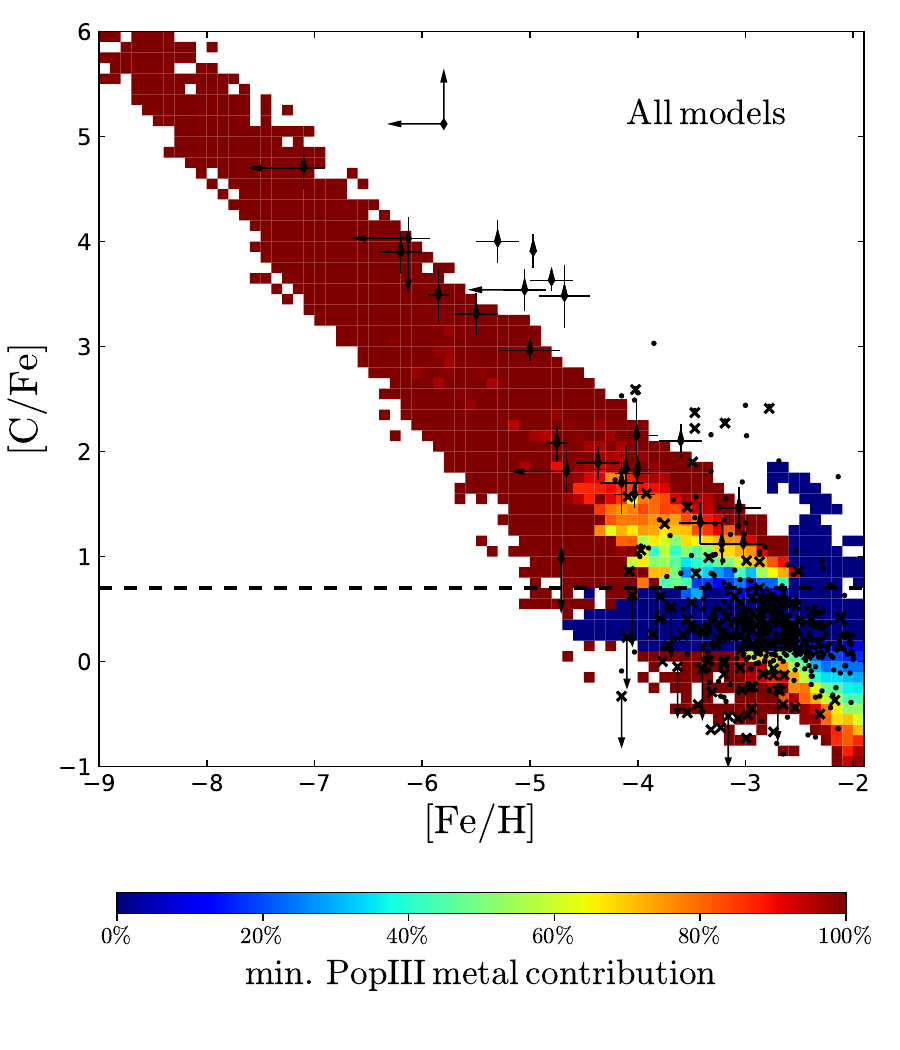}
\end{array}$
\end{center}
\caption{Left: Distribution of very metal-poor, inner-halo stars on the [Fe/H]--[C/Fe] diagram. The color denotes the probability $P_i \equiv N^i_*/N^{\rm tot}_*$ of stars to belong in each bin $i$ (averaged among all models of Table~\ref{t:t1}), where $N^i_*$ is the number of stars in the bin and $N^{\rm tot}_*$ is the total number of ${\rm [Fe/H]}\le -2$ stars. Right: Same as left panel only here the color denotes the \textit{minimum} metal fraction inherited by Pop~III ancestors for all the stars in each [Fe/H]--[C/Fe] bin. Datapoints in both panels, show the C-normal and CEMP-no stars from the samples of \citeauthor{Yong2013a} (\citeyear{Yong2013a}; $\times$ points), \citeauthor{placco2014carbon} (\citeyear{placco2014carbon}; points), and various authors (diamonds; see Section~\ref{Data}). }
\label{i:ssp}
\end{figure*}

Fig.~\ref{i:ssp} (left) shows the present-day distribution of VMP inner halo stars in the [C/Fe]-[Fe/H] diagram, averaged among all models of Table~\ref{t:t1}. Notice that the slope of the [C/Fe]-[Fe/H] relation and the region showing the highest probability (i.e., that of C-normal stars) do not strongly depend on the choice of model. However, the exact value of $P_i$ in each [C/Fe]-[Fe/H] bin is model-dependent. For example, models with a higher number of low energy SNe (higher $\alpha_e$) and/or a lower characteristic mass of the IMF, produce more stars, i.e., a higher probability, $P_i$, at $\rm[Fe/H]\lesssim-4$  (see Fig.~\ref{i:mdf}). Therefore, the average probability shown here is only a rough estimate, as $P_i$ varies across models and there is no reason to assume that each of the models in Table~\ref{t:t1} should have equal weight in the calculation of the average.

The main bulk of the observed C-normal population is in very good agreement with our model predictions, and coincides with the region predicted to have the highest density of stars. Furthermore, the sparser CEMP-no stars are also well represented by our models.
Similar to the observations, our models show a sharply decreasing [C/Fe] with increasing [Fe/H]. However, our [C/Fe]-[Fe/H] relation appears shifted towards lower [Fe/H] compared to the observed one. As a result the CEMP stars with the highest carbonicities in each [Fe/H] bin are not reproduced by our models. This is a problem faced by several other works (e.g., \citealp{Cooke2014, Komiya2020,Jeon2021}). In Section~\ref{Conclusion}, we discuss possible solutions to this discrepancy coming both from the modelling and from the observational side.

The right panel of Fig.~\ref{i:ssp}, depicts the \textit{minimum} metal fraction contributed by Pop~III stars, as a function of metallicity and carbon enhancement. In particular, the colors denote the minimum $f_Z^{\rm Pop\:III} \equiv m_Z^{\rm Pop\: III}/m_Z^{\rm tot}$ of all stars belonging to each [C/Fe]-[Fe/H] bin in our models (Table~\ref{t:t1}), where $m_Z^{\rm tot} = m_Z^{\rm Pop \: III} + m_Z^{\rm Pop \: II}$ is the total mass of metals in a star and $m_Z^{\rm Pop \: III}$ and $m_Z^{\rm Pop \: II}$ the metals' mass that it
has inherited from Pop~III and Pop~II progenitors, respectively. 
We find that all C-enhanced stars at $\rm[Fe/H]<-3$, are at least $\sim20\%$ enriched by Pop III progenitors.\footnote{Note that our model does not include binary transfer, i.e., CEMP-s stars.} As we go towards higher [C/Fe], this value increases rapidly, to $>50\%$ for stars with $\rm[C/Fe]>+1$, and to $>80\%$ for stars with $\rm[C/Fe]>+1.5$ (at the same $\rm[Fe/H]<-3$). Moreover, all stars with $\rm[C/Fe]\gtrsim +2$ and/or $\rm[Fe/H]\lesssim -4.7$ are \textit{pure} Pop~III descendants; their abundance patterns are less than 5$\%$ contaminated by Pop~II stars. 

In addition to the Pop~III enriched CEMP stars, there exists a group of C-enhanced stars that have been entirely enriched by Pop~II stars, at $\rm[Fe/H]>-2.8$ and $\rm[C/Fe]< +2$ (dark blue area in Fig.~\ref{i:ssp}). Our adopted Pop~II SN yields (\citealp{Limongi2018}) have a maximum $\rm[C/Fe]=+0.69$ at [Fe/H]$\leq-2$, and are, therefore, not able to beget VMP C-enhanced stars. Instead, we find that these (Pop~II enriched) CEMP stars are descendants of Pop~II AGB stars and can form in minihalos only after the following conditions are met: (i) SN explosions expel all, or nearly all, gas from the halo, leaving none, or only a small fraction, of the iron they produced, (ii) subsequent accretion of pristine/metal-poor gas from the IGM leads to low [Fe/H] in the ISM, (iii) previously formed AGB stars release carbon enriching the ISM to high [C/Fe] (see also \citealt{Rossi2023}).

The stars at $\rm0<[C/Fe]<+0.7$ are predominantly enriched by Pop~II progenitors. They correspond to the highest density region of Fig.~\ref{i:ssp} (left), which implies that most of the observed C-normal stars are not Pop~III star descendants. Yet, at low $\rm[C/Fe]<0$, the Pop~III metal contribution starts dominating again. This is a natural consequence of the fact that Pop II stars yield  a minimum $\rm[C/Fe]\approx0.07$ at $\rm[Fe/H]\leq-2$ \citep{Limongi2018}, while energetic Pop~III SNe can reach down to $\rm[C/Fe]\approx -1.3$  (\citealp{Heger2002, Heger2010}; see Section~\ref{Comp}).

\begin{figure*}
\begin{center}
\includegraphics[width=0.85\hsize]{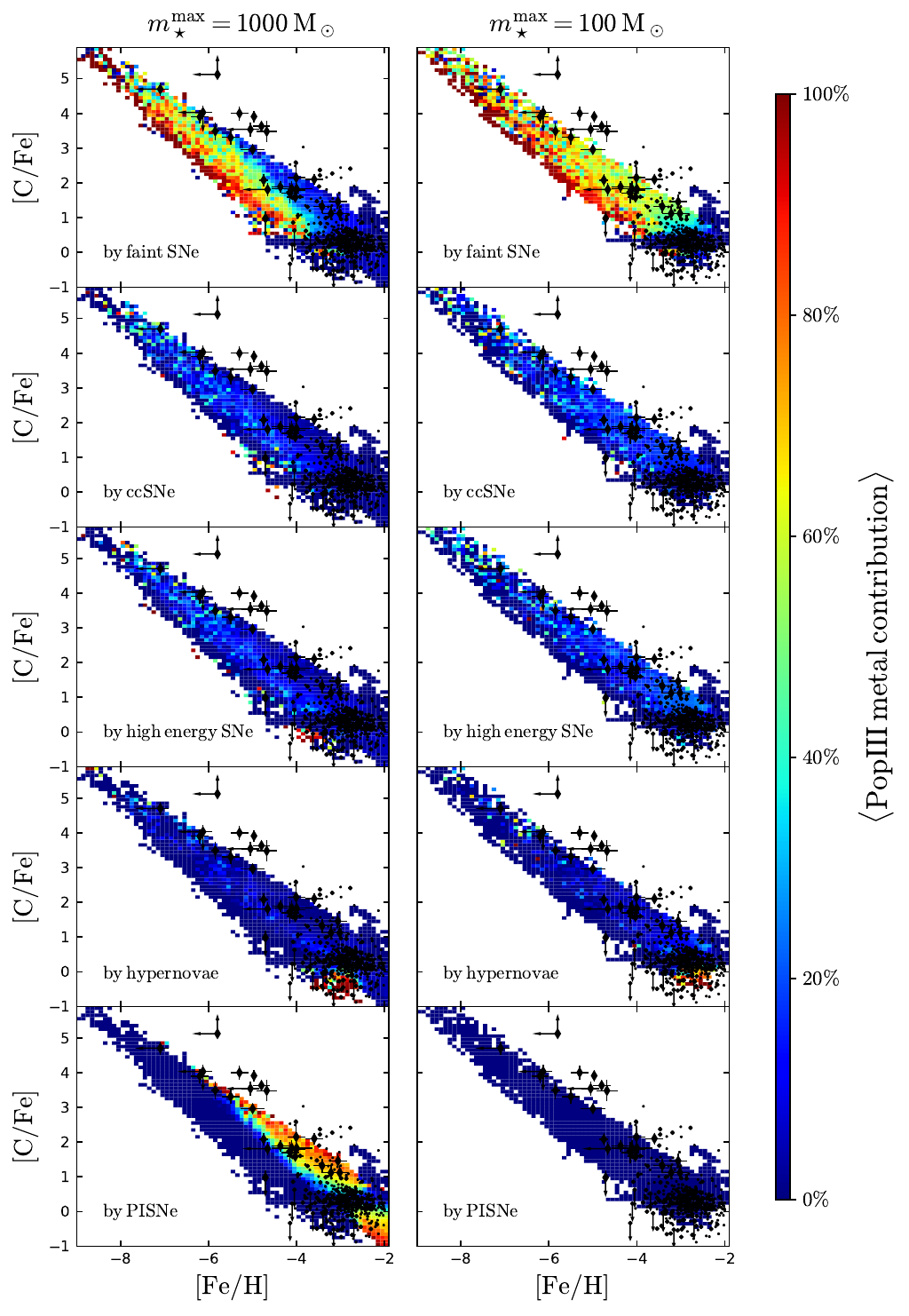}
\end{center}
\caption{Distribution of very metal-poor, inner-halo stars in the [C/Fe]--[Fe/H] diagram, with the color denoting the mean fraction of metals inherited from Pop III faint SNe, ccSNe, high energy SNe, hypernovae, and PISNe, based on the models of Table~\ref{t:t1}, with Pop III mass range $m_\star = 0.1-1000\:{\rm M_\odot}$ (left); and $m_\star = 0.1-100\:{\rm M_\odot}$, i.e., when no PISNe are allowed to form  (right). Datapoints show the observations as in Fig.~\ref{i:ssp}.}
\label{i:table}
\end{figure*}

Fig.~\ref{i:table} shows the {\it average} contribution by different Pop~III progenitors for stars in each [C/Fe]-[Fe/H] bin, for all models in Table~\ref{t:t1} with: $m_\star^{\rm max}=1000\:{\rm M_\odot}$ (left); and  $m_\star^{\rm max}  = 100\:{\rm M_\odot}$, i.e.,~when no PISNe are allowed to form (right).
We find that at fixed [C/Fe], the surviving CEMP stars with the lowest [Fe/H] have been enriched mostly by faint SNe. Instead, PISNe enrichment dominates the pollution of the most [Fe/H]-rich CEMP stars. Notice that yields from PISNe with $m_\star = (140-150)\:{\rm M_\odot}$ can reach significantly higher [C/Fe] than many non-PISNe, e.g. \citealp{Heger2010, Nomoto2013}. In the absence of PISNe, however, all CEMP stars (except the Pop~II AGB-descendants at $\rm[Fe/H]>-2.8$) are on average $>30\%$ enriched by faint SNe. Compared to faint SNe and PISNe, the overall contribution of ccSNe, high energy SNe and hypernovae to the surviving stars is not as prominent. This is due to the fact that the EDFs of our preferred models (Table~\ref{t:t1}) are skewed towards low explosion energies --- high $\alpha_e$ exponent --- and hence produce much fewer SNe of high energies. Nonetheless, there appears a region in the diagram (at $\rm[C/Fe]<0$ and $\rm[Fe/H]\lesssim-2.5$) where stars are predominantly enriched by primordial hypernovae. For comparison, in Appendix~\ref{appendix1} we show the results of a model in which all types of Pop~III SNe are equally probable. 

The average contribution of any SN type, at a given [C/Fe]--[Fe/H] bin, varies depending on the assumed EDF. However, the qualitative trends described above do not: we find that long-lived descendants of each kind of Pop~III SNe always occupy the same regions on the [C/Fe]--[Fe/H] diagram. In particular, certain [C/Fe]--[Fe/H] combinations can {\it only} be produced by an enrichment of a specific type of Pop~III SN. Hypernovae descendants predominantly populate a well defined region at $\rm[C/Fe]< 0$ and $\rm[Fe/H]\lesssim-2.5$, while the $\rm[C/Fe]\lesssim-0.5$ area is only populated with PISNe descendants. Thus, without hypernovae or PISNe these areas are not represented in our models (compare, e.g., the bottom panels in Fig.~\ref{i:table}).

\section{The impact of stellar mixing}
\label{mixing}

The convective mixing between stellar layers can influence quite strongly the chemical signature of Pop~III SNe. When mixing precedes fallback, heavier nuclei that would not have been ejected otherwise can escape into the ISM. In the previous Section, we adopted a uniform distribution for the mixing parameter, $f_{\rm mix}$ of the \citet{Heger2010} yields (see Section~\ref{sec:yields_mixing}). Here, we explore how varying $f_{\rm mix}$ impacts our model's predictions.

\begin{figure*}
\begin{center}
\includegraphics[width=0.9\hsize]{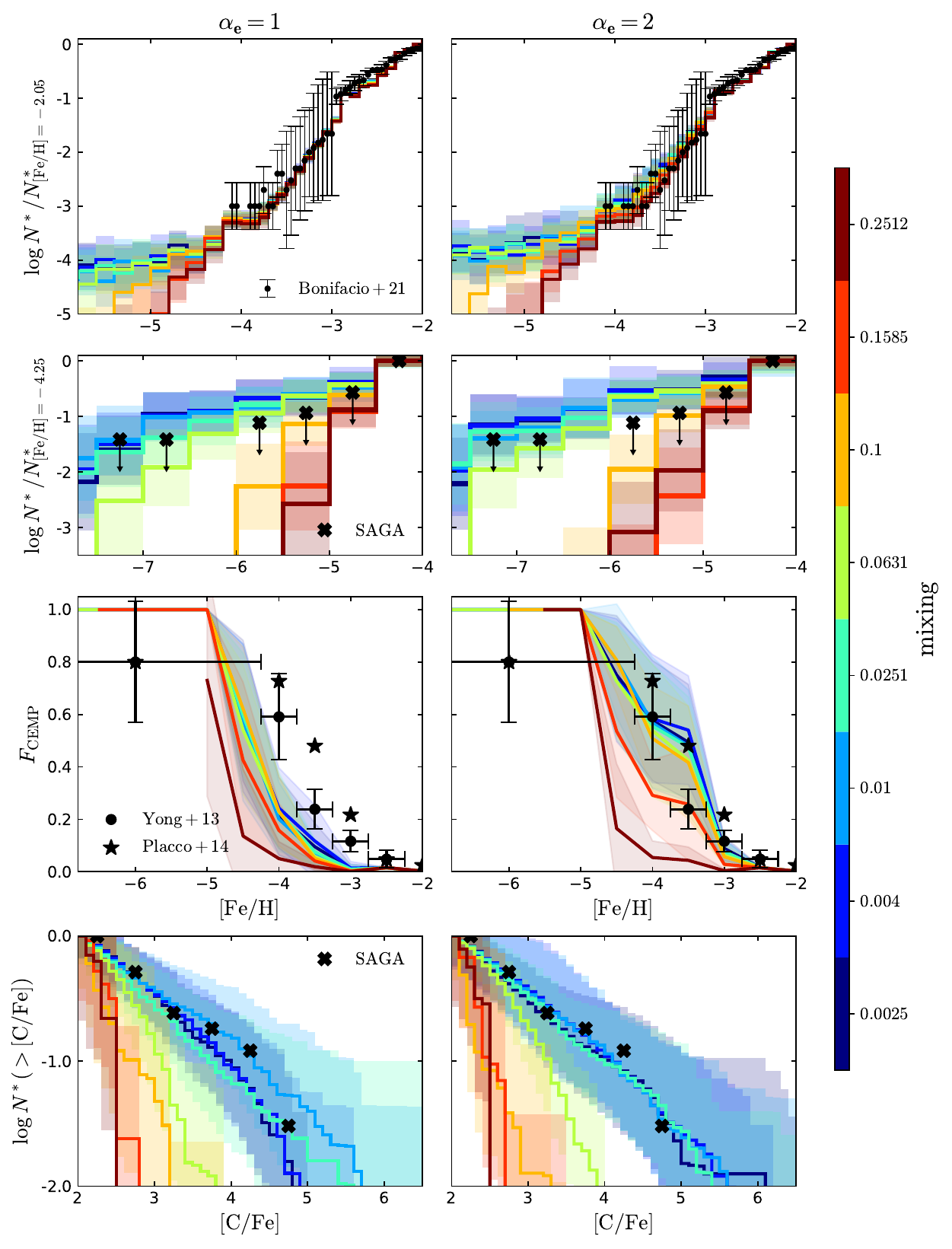}
\end{center}
\caption{Predicted metallicity distribution function of inner halo stars ($7\:{\rm kpc}<R_{\rm gal}<20\:$kpc), normalized at $\rm[Fe/H]=-2.05$ (top row), and $\rm[Fe/H]=-4.25$ (second row), in comparison with observations from \citet{Bonifacio2021} and the SAGA database (see Section~\ref{Data} for details). The third row shows the predicted fraction of CEMP stars ($\rm[C/Fe]\geq+0.7$) in each [Fe/H] bin, in comparison to the observations by \citeauthor{Yong2013a} (\citeyear{Yong2013a}; points with errorbars) and \citeauthor{placco2014carbon} (\citeyear{placco2014carbon}; stars). The bottom row shows the cumulative [C/Fe] distribution function (CDF) of VMP inner halo stars, normalized at $\rm[C/Fe]=+2$, in comparison with observations from the SAGA database. Colors in all panels denote the mixing parameter adopted for $m_\star=10-100\:{\rm M_\odot}$ Pop~III stars in each run, as indicated by the colorbar. Lines and shaded areas represent the mean and standard deviation of 100 runs.
The left and right panels show the model results assuming an EDF for Pop~III SNe (Eq.~\ref{e:EDF}) with exponent $\alpha_e=1$ and $\alpha_e=2$, respectively. In both cases higher mixing for Pop~III stars results in steeper MDFs and CDFs and lower CEMP fractions.}
\label{i:mix}
\end{figure*}

The left and right columns of Fig.~\ref{i:mix}, show our results as a function of $f_{\rm mix}$, assuming an EDF given by Eq.~\ref{e:EDF} with $\alpha_e = 1$ and $\alpha_e = 2$, respectively. Both cases assume a Pop~III IMF of the form of \citeauthor{larson1998early} (\citeyear{larson1998early}; Eq.~\ref{e:Larson}) with $m_{\rm ch} = 10\:{\rm M_\odot}$ and $m^{\rm max}_{\star} = 1000\:{M_\odot}$, i.e., the second IMF considered in Section~\ref{IMF}.

Similar to the case of varying the Pop~III EDF and IMF (Section~\ref{IMF}), the predicted MDFs at [Fe/H] > -3 remain unaffected by the assumed stellar mixing and in agreement with the one by \citeauthor{Bonifacio2021} (\citeyear{Bonifacio2021}; 1st row of Fig.~\ref{i:mix}). At $\rm[Fe/H]<-3$, the MDFs resulting from different models start to differentiate and this difference becomes pronounced at [Fe/H]$\leq-4$; for a given EDF, lower mixing levels produce flatter MDFs. The comparison with the SAGA MDF is crucial to potentially discard models (2nd row of Fig.~\ref{i:mix}). Models with $f_{\rm mix}\leq 0.0631$ produce MDFs that lie above the observational values in several metallicity bins. Still we cannot completely discard them since they are in agreement with the observations within errors. On the other hand, higher mixing levels ($f_{\rm mix}\geq 0.1$) result in more heavy elements, such as iron, being released into the ISM. The chemical enrichment in the first minihalos proceeds faster giving rise to steeper MDFs that underpredict the number of hyper metal poor stars by more than 1-2\,dex. 

\begin{figure*}
\begin{center}
    \includegraphics[width=0.94\hsize]{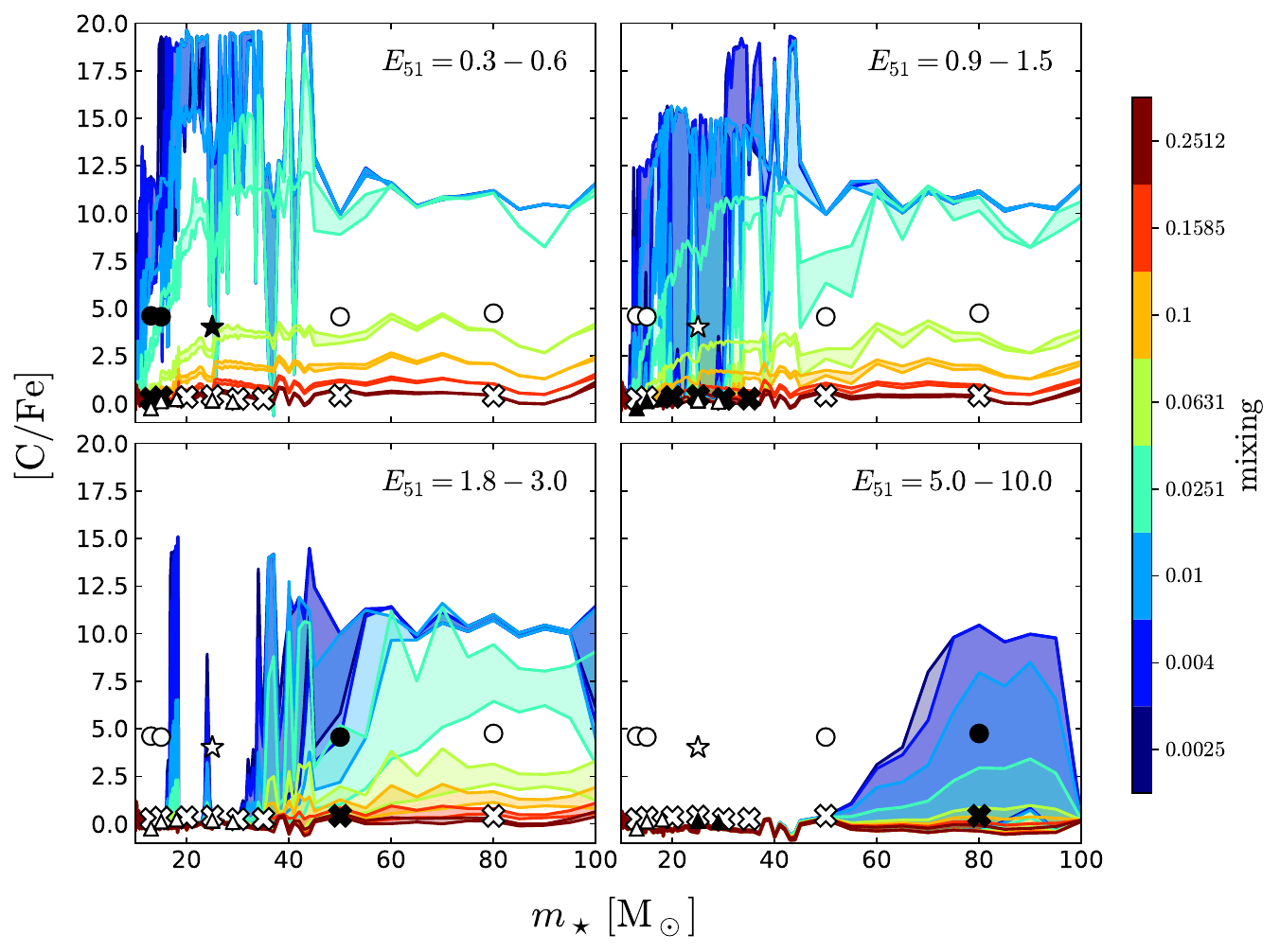}
\end{center}
\caption{Comparison of the \citet{Heger2010} [C/Fe] yields for Pop III faint SNe (${E_{51}=0.3-0.6}$; top-left), ccSNe (${E_{51}=0.9-1.5}$; top-right), high energy SNe (${E_{51}=1.8-3.0}$; bottom-left) and hypernovae (${E_{51}=5-10}$; bottom-right panel) with the [C/Fe] yield of zero-metallicity stars by \citeauthor{Iwamoto2005} (\citeyear{Iwamoto2005}; stars), \citeauthor{Tominaga2007} (\citeyear{Tominaga2007}; triangles),  \citeauthor{Marassi2014} (\citeyear{Marassi2014}; circles) and \citeauthor{Limongi2012} (\citeyear{Limongi2012}; X symbols). 
Colors denote the mixing level given by \citet{Heger2010}, as indicated by the colorbar. For each mixing level, the top (bottom) coloroud solid line corresponds to the lowest (highest) explosion energy shown in each panel, with the area between the two lines shaded. Filled black datapoints show [C/Fe] yields of Pop III stars with explosion energies within (or close) to the range shown in each panel, while empty datapoints show SNe yields of different energies.}
\label{i:MixComp}
\end{figure*}

The third row of Fig.~\ref{i:mix} shows the fraction of CEMP-no stars, $F_{\rm CEMP}$ (Eq.~\ref{e:F_CEMP}), in the inner-halo, as predicted by the different models. We find that mixing levels $f_{\rm mix} \leq 0.1$ yield almost identical $F_{\rm CEMP}$ in both models. Those lie well below the observations in the $\alpha_e=1$ model, but are in agreement with the observations in the $\alpha_e =2$ model, due to the lower number of high energy SNe and hypernovae there (7$\%$ for $\alpha_e=2$ compared to 43$\%$ for $\alpha_e=1$).
The two highest mixing levels, $f_{\rm mix} = $0.1855 and 0.2512, underestimate the CEMP fractions in both models and do not produce any stars at $\rm[Fe/H]<-6.5$. We also notice that the $F_{\rm CEMP}$ dependence on mixing is weaker for the $\alpha_e=1$ model, which generates more energetic SNe. Indeed, the higher the explosion energy, or equivalently the smaller the fallback, the weaker the effect of mixing on stellar ejecta\footnote{In the limit where the fallback is zero, the stellar yields are independent of mixing.} (see Fig.~\ref{i:MixComp}).

The $F_{\rm CEMP}$--[Fe/H] relation can convey only limited information, since CEMP is a binary classification -- a star either has $\rm[C/Fe]>+0.7$ or not. The CDF, instead, can be more informative. The bottom row of Fig.~\ref{i:mix} shows the cumulative CDF for all inner halo stars with $\rm[Fe/H]<-2$ and $\rm[C/Fe]>+2$ in each model. Here, the effect of mixing is pronounced in both the $\alpha_e=1$ and the $\alpha_e=2$ model. In both cases, models with $f_{\rm mix}\geq 0.0631$ predict too few stars with high carbonicities ($\rm[C/Fe]\gtrsim+3$). Instead, the CDFs for lower $f_{\rm mix}$ are in good agreement with the observations for $\alpha_e=1$ and even better for $\alpha_e=2$. 

Regardless of the degeneracy between the stellar mixing and the explosion energy in the yielded abundance ratios --increasing either the mixing or the explosion energy yields lower [C/Fe]-- we find that when the fraction of energetic SNe is significant (e.g., when $\alpha_e=1$) even the lowest mixing level is inconsistent with the observations. Similarly, the highest mixing levels fail to reproduce the observations even when the EDF is dominated by faint SNe. More specifically, inspection of the second and fourth row of Fig.~\ref{i:mix} reveals that a typical $f_{\rm mix}\leq0.0631$ is favoured by our model. This result is not at odds with our adoption of a uniform mixing distribution in Sections~\ref{IMF} and \ref{POPIII}, since only 3 out of the 14 mixing levels of \citep{Heger2010} are above $f_{\rm mix}=0.0631$. 
We should note that \citet{Heger2010} reached similar conclusions. In particular, they found that regardless of the assumed Pop~III IMF and explosion energy, an $f_{\rm mix}\leq 0.0631$ provides the best fit to the abundance patterns of the C-enhanced stars HE1327-2326 \citep{Frebel2005} and HE0107-5240 \citep{Christlieb2002}. 

It would be valuable to generalise this result to other stellar evolutionary models. However, this endeavor is no easy feat, first, due to the fact that the mixing prescription in \citet{Heger2010} is artificial and not described using physical principles, and second, due to the many differences in the physical and numerical assumptions employed by different groups. Nevertheless, one can get a crude idea by comparing the yielded abundance ratios obtained in different studies to their adopted parameter values.

Fig.~\ref{i:MixComp} compares the [C/Fe] yields provided by \citet{Heger2010} for Pop~III SNe as a function of mixing and explosion energy, to those computed for Pop~III SNe by \citet{Iwamoto2005}, \citet{Tominaga2007}, \citet{Marassi2014} and \citet{Limongi2012}. In the first three works, mixing is a free parameter like in the \citet{Heger2010} models but is parametrized in terms of a minimum and maximum mass coordinates within which the mass of each element is uniformly mixed. Those two together with the "mass cut" parameter, i.e., the mass coordinate below which all material falls back onto the central remnant, are calibrated\footnote{One can find the adopted parameter values in the respective papers.} to reproduce the abundance pattern of the stars: (i) HE1327-2326 and HE0107-5240 in the case of \citet{Iwamoto2005} (ii) the CEMP star SMSSJ031300 observed by \citet{Keller2014} in the case of \citet{Marassi2014}; and (iii) the average abundance pattern of four C-normal EMP stars \citet{Cayrel2004} in \citet{Tominaga2007}. Naturally, a more extended mixing region at fixed mass cut, or a lower mass cut at fixed mixing, would result in lower yielded [C/Fe]. In \citet{Limongi2012}, stellar mixing is not artificial but it is coupled to nuclear burning and only the mass cut (or equivalently the explosion energy) is calibrated to reproduce the average abundance pattern of the C-normal stellar sample of \citet{Cayrel2004}. 

Since the free parameters in the above models have been fitted to reproduce specific stars, they give a quasi-constant [C/Fe], regardless of the assumed stellar mass; the \citet{Iwamoto2005} and \citet{Marassi2014} yields that have been calibrated to reproduce CEMP stars, lie above our typical $f_{\rm mix}$=0.0631 level in all energy bins, while the ones of \citet{Tominaga2007} and \citet{Limongi2012} that have been calibrated to reproduce C-normal stars are consistent with the highest mixing levels and/or the highest explosion energies of \citet{Heger2010}.

\section{Discussion}
\label{Discussion}
We have developed a new SAM of galaxy formation, named {\sc NEFERTITI} ({\it NEar-FiEld cosmology: Re-Tracing Invisible TImes}), and combined it with a cosmological N-body simulation of a MW analogue, to shed light on the properties of the first Pop~III SNe, and in particular their energy distribution function, EDF, parameterized in our model with $\alpha_e$ (Eq.~\ref{e:EDF} and Fig. \ref{i:EDF}). {\sc NEFERTITI}  follows the formation and evolution of individual Pop III and Pop II stars, and considers, for the first time in a SAM for the MW formation, 
the contribution of Pop III SNe with different masses, stellar mixing and explosion energies. Subsequently, we have investigated how varying these Pop III stellar parameters affects the properties of the surviving VMP stars in the galactic halo. 

\subsection{Comparison with previous results and model limitations}
\label{Comp}

For a given Pop III IMF, we find that a higher contribution of low energy SNe (higher $\alpha_e$) results in a greater CEMP-no fraction, and a flatter MDF and CDF for stars with $\rm[Fe/H]<-2$. In particular, for a \citet{larson1998early} IMF with characteristic mass $m_{\rm ch}=1\:{\rm M_\odot}$ and range $m_\star = [0.1, 1000]\:{\rm M_\odot}$, we reproduce the observed CEMP fractions by \citet{Yong2013a} and \citet{Placco2021} if 40\%-80\% of Pop~III stars with $10\leq m_\star/{\rm M_\odot} \leq100$ explode as faint SNe and only 20$\%$-2$\%$ as hypernovae (and the rest with intermediate energies; Fig.~\ref{i:EDF}). When we increase $m_{\rm ch}$ to $10\:{\rm M_\odot}$, the data are best fitted with 63$\%$-90$\%$ of faint SNe, while for $m_{\rm ch}=100\:{\rm M_\odot}$ we always underpredict the fraction of CEMP stars, due to the high number of primordial PISNe (Fig.~\ref{i:cemp}). The effects of the Pop III properties on the halo MDF are only prominent at $\rm[Fe/H]\lesssim-3$, yet all models are in agreement with the current observations within errors (Figs~\ref{i:mdf} and \ref{i:mdf2}). However, this comparison is somewhat hampered by limitations of the data (see the next subsection).

These results are not easily compared to previous works due to the different assumptions taken (e.g., the Pop III IMF), and the fact that, until now, none have considered simultaneously the contribution of Pop III SNe of all different energies and Pop II stars. 
However, we report that \citet{Hartwig2018a} find a best matching to the \citet{placco2014carbon} CEMP fractions by assuming that 40\% of Pop~III stars explode as faint SNe and the rest as normal ccSNe, but without including high energy SNe, hypernovae and PISNe with $m_\star > 150\:{\rm M_\odot}$,  all of which would lower $F_{\rm CEMP}$. 
Contrarily, the simple minihalo model for Pop~III star enrichment by \citealp{Cooke2014} manages to reproduce the observed $F_{\rm CEMP}$ even when 100$\%$ of Pop III SNe explode with high energy, when using a flat IMF in the range 10-100$\: {\rm M_\odot}$ (hence without including PISNe). This overestimation of the CEMP fraction relative to our model is to be expected, since \citet{Cooke2014} do not consider the contribution from Pop~II stars which, according to our simulation, dominates the chemical enrichment at $\rm[C/Fe]<+0.7$ (see also Vanni et al. 2023).

One of our key findings is that all VMP stars with subsolar [C/Fe] are predominantly imprinted by Pop~III hypernovae and/or PISNe, regardless of our model assumptions (Fig.~\ref{i:table}). This result stems from the fact that our adopted Pop~II metal yields \citep{Limongi2018} reach a minimum $\rm[C/Fe]= 0.07$ at $\rm[Fe/H] \leq -2$, whereas Pop~III hypernovae and PISNe can reach [C/Fe] as low as $\approx-0.9$ and $\approx - 1.3$, respectively \citep{Heger2002, Heger2010}. The uncertainties associated with Pop~II yields are crucial in this regard. \citet{Ritter2018} similarly estimate a yielded $\rm[C/Fe]>0$ for massive stars with $\rm [Fe/H]<-2$, and \citet{Kobayashi2006} and \citet{Nomoto2013} find a yielded $\rm[C/Fe]<-0.1$ only at $\rm [Fe/H]<-3.5$ (see Fig.~5 of \citealp{Liang2023}). Yet, \citet{Woosley1995} suggest Pop~II [C/Fe] yields reaching down to $\approx-0.5$. Therefore, only if we adopted the \citet{Woosley1995} yields, we would anticipate a higher contribution of Pop~II stars at subsolar [C/Fe]. It should be noted that our model does not include SN type~Ia that yield $\rm[C/Fe]<-1$ \citep{Thielemann1986,Iwamoto1999, Seitenzahl2013}. However, we expect their contribution at such low metallicites to be minimal \citep[e.g.][]{salvadori15}. Type~Ia SNe have a typical delay time of $0.1-1$~Gyr (see \citealp{Chen2021} and references therein), while the majority of our SF minihalos reach $\rm[Fe/H]>-2$ within 0.1~Gyr of their formation. In any case, SNe type Ia descendants can be distinguished from those of Pop~III hypernovae and PISNe by comparing their complete abundance patterns (see, e.g., Fig.~8 in \citealp{Nomoto2013} and Fig.~12 in \citealp{salvadori2019probing}).

Besides the success of our model in reproducing the MDF, the CDF and the fraction of CEMP stars at [Fe/H]$\leq-2$, we find that our predicted ${\rm [C/Fe]-[Fe/H]}$ relation lies at the lower side
of the observations (Fig.~\ref{i:ssp}).
This discrepancy has been reported by several previous works, even though the [C/Fe]-[Fe/H] stellar distribution depends on the particularities of each assumed model.\footnote{For example one can easily infer from equations~\ref{e:sfr} and \ref{e:Z_ISM} that at fixed [C/Fe], [Fe/H]$\propto \epsilon_{\rm SF}$.} \citet{Cooke2014}, who investigate early chemical enrichment by Pop III stars in isolated minihalos, find that they cannot reproduce the highest [C/Fe] observed in CEMP-no stars. 
Using a model calibrated on the UFD B{\"o}otes~I, \citet{Rossi2023} find that true Pop~III descendants have\footnote{A(C)$\equiv {\rm log(N_{C}/N_H)}+12$} $\rm A(C)<6$, similar to us: the upper envelope of our [C/Fe]--[Fe/H] relation corresponds to $\rm A(C)\sim 6-6.5$ at $\rm[Fe/H]\leq-4.5$. 
Yet, they predict the formation of Pop II AGB descendants with high C-abundances (A(C)$\sim$7-7.5) even at $\rm[Fe/H]<-4$.
As explained in Section~\ref{POPIII}, such CEMP stars only form after one or more SN explosions blow out $\sim$all gas from within a halo. This process removes the iron rich signature of Pop II SNe, allowing previously formed AGB stars to enrich the newly accreted, nearly pristine gas to high [C/Fe]. 
We find that this condition is satisfied in our model only at $z<8$ when the IGM has already been enriched to $\rm[Fe/H]>-3$ (Fig.~\ref{i:ssp})\footnote{The mass loading factor $\eta \equiv \dot{M}_{\rm gas,ej}/{\rm SFR} \propto 1/u_{\rm esc}^2$ (Eq.~\ref{e:Mej}) is a decreasing function of redshift \citep{barkana2001beginning}, therefore a complete blown-out of the gas at fixed $M_{\rm vir}$ occurs more easily at low $z$.}. However, we must note that our DM simulation does not resolve halos with $M_{\rm vir} < 10^7 \: {\rm M_\odot}$. Hydrodynamic simulations, instead, suggest that the minimum mass of the first SF minihalos can range between $10^{5.5}-10^{7.5}\:{\rm M_\odot}$ depending on the relative velocity between baryons and DM (see \citealt{Schauer2019} and references therein). Including lighter minihalos in our simulation could allow the formation of AGB-descendant CEMP stars at lower metallicities. In such case, one must make sure that their abundances in s-process elements, inherited by their AGB progenitors, are in accordance with observations.

Another factor that can affect our results is the assumption of the instantaneous mixing approximation.\footnote{Notice that under the instantaneous mixing approximation, including lighter minihalos would move our [C/Fe]-[Fe/H] relation leftwards, i.e., further away from the observations, since metal enriched gas escapes more easily from less massive halos, thereby lowering [Fe/H] while keeping [C/Fe] (which depends only on the assumed metal yields) $\sim$constant.} In reality, SNe ejecta may only mix with a fraction of the available cold gas (see \citealp{salvadori2019probing} and \citealp{Magg2020} for an estimate of the dilution factor). The resulting abundances in the SF clouds are higher, due to less dilution, and may even differ from the gross yields of the SNe (towards higher [C/Fe]; \citealt{Ritter2015}). However, the minimum dilution mass can not be arbitrarily small but is limited by the mass enclosed within the final size of the SN remnant \citep{Magg2020}. Models that take into account realistic prescriptions for the diffusion of SNe ejecta still find it challenging to reproduce CEMP stars with A(C)$\sim$7-7.5 (\citealp{Chiaki2020, Jeon2021}, Vanni et al. in prep).
\citet{Komiya2020} find that only an extremely inefficient mixing of SN yields can reproduce the highest [C/Fe] CEMP-no stars, but this results in an inconsistent metallicity distribution function. They concluded that binary mass transfer from AGB stars is neccessary to explain the [C/Fe] abundances of $\rm[Fe/H]<-4$ stars. 
\citet{Sarmento2019} do manage to reach $\rm A(C)\gtrsim7.5$ using an Pop III IMF with $m_{\rm ch}=60-120\:{\rm M_\odot}$ and $m_\star = [20-120]\:{\rm M_\odot}$ (the range yielding the highest [C/Fe]), but do not report on their predicted MDF and CEMP fraction. We find that adopting this IMF in our model results in a too flat MDF at $\rm[Fe/H]<-4$, inconsistent with the SAGA observations. 

Finally we must note that our results could also depend on the adopted merger tree. \citet{Chen2023}, for example, find that the predicted MDFs in different MW-like analogues can differ by $\sim1\:$dex at $\rm [Fe/H]=-4$. We plan to explore the level of this dependence in a future work. 

\subsection{Key observables and their intrinsic uncertainties}

We have shown that we can constrain the properties of primordial SNe by comparing our model predictions to observations of VMP stars in the Galactic halo. In particular, we find that both the mixing and the explosion energies as well as the IMF of Pop III stars have a strong impact on the present day CEMP-no fraction, the MDF and the CDF. 

Stellar mixing affects strongly the halo MDF at $\rm[Fe/H]<-4$, and the CDF at $\rm[C/Fe]>+2$ (Fig.~\ref{i:mix}), where the sample of high-resolution follow-up observations can be deemed unbiased. Instead, the effect of adopting different Pop III IMFs and EDFs appears more prominent when we consider a broader metallicity range, i.e. a MDF extending from the lowest [Fe/H] up to $\rm[Fe/H]>-3$ (Figs~\ref{i:mdf} and \ref{i:mdf2}), and a CDF extending from the highest carbonicities down to $\rm[C/Fe]<+1$ (Figs~\ref{i:cdf2} and \ref{i:cdf}). Unfortunately, the currently available observations at these abundance ranges are incomplete, and follow-up is biased towards the lowest metallicities and highest carbonicities.

In addition, all the aforementioned observables suffer from large observational errors. \citet{Arentsen2022} report that there are significant systematic differences in the carbon abundances among various surveys of Galactic halo stars that can translate to more than 50$\%$ differences in the estimated CEMP fractions (see their Fig.~1). Similar uncertainties apply to the determination of [Fe/H] and the MDF (see e.g., Fig.~12 of  \citealp{youakim2020pristine}). These systematics can arise from different resolution and pipeline approaches, different assumptions in the employed synthetic grids, and/or comparison of stars in different evolutionary phases. 

An additional source of large systematic errors comes from the simplifying assumption of one-dimensional (1D), local thermodynamic equilibrium (LTE) hydrostatic model atmospheres that is used in standard spectroscopic abundance analyses. Accounting for 3D non-LTE effects has been found to lower [C/Fe] estimates by as much as $\sim$1 dex while  raising [Fe/H] by $<$0.15 dex \citep{Collet2006,Amarsi2019a, Amarsi2019b, Norris2019}. Naturally, this  has a dramatic effect on the fraction of CEMP stars; \citet{Norris2019} found that after applying 3D non-LTE corrections, the \citet{Yong2013a} CEMP-no fraction at $-4.5\leq$[Fe/H]$\leq-3$ is reduced by $\sim60\%$ while the number of CEMP-no stars in the \citet{Yoon2016} sample decreases by $\sim73\%$. Correcting for 3D non-LTE effects will also move the observed [C/Fe]-[Fe/H] stellar distribution downwards and, perhaps, resolve the discrepancy with our predictions (Fig.~\ref{i:ssp}).

Finally, several CEMP stars with $\rm A(C)>6.5$ (and all of them at $\rm [Fe/H]<-4.5$) that are not reproduced by our model, have either no Ba measurements or have only upper limits for barium enhancement at [Ba/Fe]>0.6. If a high Ba enhancement is confirmed for those stars in the future, then their high carbonicities could be explained by enrichment from a Pop II AGB progenitor \citep{Rossi2023} or mass transfer from a Pop III/II AGB companion \citep{Komiya2020}.

\section{Conclusions and future outlook}
\label{Conclusion}

For the first time, we explore the energy distribution function, EDF, of the first SNe in the context of a cosmological galaxy formation model of a MW-analogue. Our model follows the formation and evolution of individual Pop III stars, which is uniquely determined by their initial mass, stellar mixing and explosion energy. Their contribution in the chemical enrichment of their host minihalos is imprinted in the present day properties of very metal-poor galactic halo stars, such as their MDF, CDF and CEMP fractions. We draw the following main conclusions:

\begin{enumerate}
\item{{\it Pop~III Energy Distribution Function.} The fraction of CEMP stars, $F_{\mathrm{CEMP}}$, is highly sensitive to the primordial EDF, especially at $\rm[Fe/H]<-3$. Assuming an EDF of the form: $dE/dN \propto E^{-\alpha_e}$, we find that we can reproduce the observed CEMP fractions for $\alpha_e\sim1-2.5$, depending on the adopted IMF for Pop III stars (Fig.~\ref{i:cemp} and Table~\ref{t:t1}). This value corresponds to a $\sim40-90\%$ probability for Pop III stars with $m_\star=10-100\:{\rm M_\odot}$ to explore as faint SNe, and a $20-0.5\%$ probability for them to explode as hypernovae (intermediate energy SNe have intermediate probabilities; Fig.~\ref{i:EDF}). The effect of the Pop III EDF (and of their IMF) on the halo MDF is only prominent at $\rm[Fe/H]\lesssim-3$ but there the observational uncertainties are so large that they render any comparison inconclusive (Figs~\ref{i:mdf} and \ref{i:mdf2}).}
\item{{\it Pop~III Inital Mass Function.} A top-heavy primordial IMF (with characteristic mass $m_{\rm ch}=100\:{\rm M_\odot}$ in the range 0.1-1000$\:{\rm M_\odot}$) is disfavoured, as it underestimates the CEMP fraction and results in a too steep CDF, even if all Pop~III stars with $m_\star=10-100\:{\rm M_\odot}$ explode as faint SNe (Figs~\ref{i:cemp} and \ref{i:cdf2}).}

\item{{\it Pop~III stellar mixing.} At a given EDF and IMF, lower mixing for Pop III stars results in a flatter MDF, higher CEMP fractions and a CDF skewed towards higher [C/Fe]. We find that typically very low mixing ($f_{\rm mix}\leq0.0631$ as provided by \citealp{Heger2010}) is required to reproduce the observations (Fig.~\ref{i:mix} and~\ref{i:MixComp}).}

\item{{\it Pop~II descendants.} The great majority of very metal-poor stars lie at $0< {\rm [C/Fe]}< +0.7$, i.e. they are C-normal. We predict that these stars have been predominantly polluted by normal Pop~II SNe, in agreement with recent studies investigating the abundance patterns of C-normal stars and their small star-to-star scatter (Vanni et al. 2023). In addition, we find a population of CEMP stars at $\rm [Fe/H]\gtrsim -2.8$, which were born from gas enriched by Pop~II AGB stars.}
\item{{\it Pop~III descendants.}
Regardless of the assumed model, all CEMP stars at $\rm[Fe/H]\lesssim -2.8$ have been enriched to $>20\%$ by Pop III progenitors. This value increases to $>95\%$ at $\rm[C/Fe]\gtrsim +2$ (Fig.~\ref{i:ssp}). At fixed [C/Fe], CEMP stars with the lowest metallicities are faint SNe descendants, while as we move to higher [Fe/H] the contribution of higher energy Pop III SNe prevails.
According to our results, very metal-poor stars with $\rm[C/Fe]\lesssim 0$ are predominantly imprinted by primordial hypernovae (at $\rm[Fe/H]\lesssim-2.5$) and PISNe (at $\rm [Fe/H]\gtrsim-2.5$; Fig.~\ref{i:table}).}

\end{enumerate}

 We have demonstrated that the Pop III EDF can be equally important to their IMF in shaping the abundances of EMP halo stars. We find that only EDFs that are weighted towards low explosion energies combined with bottom heavy IMFs (even if they extend to 1000$\:{\rm M_\odot}$) can reproduce simultaneously the MDF, the CDF and the fraction of CEMP stars in the Galactic halo. However, this comparison alone does not allow a tighter constraint on the Pop III IMF, mixing and EDF due to the degeneracies between them and, most importantly, to the large uncertainties associated with the observed relations.

 We have shown that, regardless of the assumed model, the descendants of each type of primordial SNe, always appear at specific regions in the [C/Fe]--[Fe/H] diagram. However, their prevalence there varies depending on the IMF and EDF of Pop III stars (Figs~\ref{i:table} and \ref{i:PLEGMAalleq}). In a following study, we will quantify this variation and compare with the observed fractions of confirmed Pop III-enriched stars. In addition, we plan to follow additional chemical elements, or even full abundance patters (e.g., Vanni et al. in prep). This way, we will gain further insight into the properties of primordial SNe and potentially break the aforementioned degeneracies. 

 Additional constraints from forthcoming large spectroscopic surveys, such as WEAVE \citep{dalton2016weave} and 4MOST \citep{Christlieb2019}, as well as surveys dedicated to identifying Pop III descendants (e.g., \citealp{Aguado2023}) and complementary studies of high-z gaseous absorption systems imprinted by the first stellar generations (e.g. Saccardi et al. 2023), will greatly boost our efforts to unveil the nature of the first SNe.

\section*{Acknowledgements}

The authors acknowledge support from the ERC Starting Grant NEFERTITI H2020/808240.

\section*{Data Availability}

The data underlying this article will be shared on reasonable request to the corresponding author.



\bibliographystyle{mnras}
\bibliography{EDF} 



\appendix

\section{The extended CDF}
\label{appendix1}

\begin{figure*}
\begin{center}
\includegraphics[width=0.75\hsize]{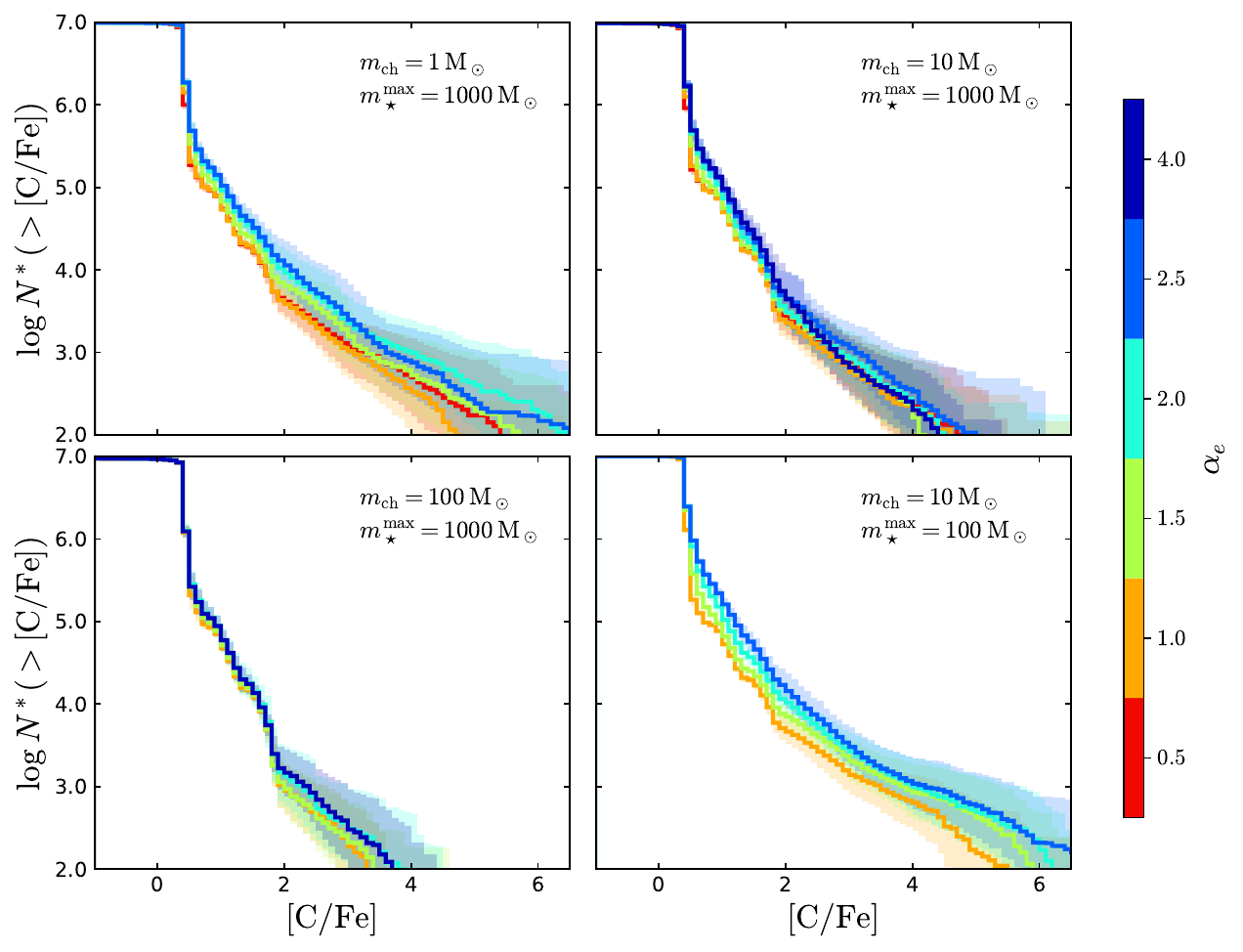} 
\end{center}
\caption{Same as Fig.~\ref{i:cdf2} but extending down to [C/Fe]=-1.}
\label{i:cdf}
\end{figure*}

Fig.~\ref{i:cdf} shows the cumulative CDF of inner halo stars with $\rm[Fe/H]\leq-2$ for the models of Section~\ref{IMF}, extending down to $\rm[C/Fe]=-1$. The total number of stars changes slightly between models, from $\sim$9.500.000 for $[m_{\rm ch}, m_\star^{\rm max}] = [100,1000]\:{\rm M_\odot}$, i.e.,~the model predicting the highest number of PISNe, to $\sim$10.000.000 for $[m_{\rm ch}, m_\star^{\rm max}] = [10,100]\:{\rm M_\odot}$, i.e.~the model with no PISNe. Yet, the slope of the CDF shows significant variation with the IMF of Pop III stars; for $m_{\rm ch}=100\:{\rm M_\odot}$ there are less than 70 stars with $\rm [C/Fe]>+4$, while for $m_\star^{\rm max}=100\:{\rm M_\odot}$ there are more than 600 stars. In addition, one can see a clear dependence of the CDFs on the Pop III EDF that becomes more prominent at high [C/Fe]; the lower the $\alpha_e$ parameter, i.e., the more the high energy SNe and hypernovae, the steeper the CDF at fixed IMF. This trend is obscured when the CDFs are normalized to $\rm[C/Fe]=+2$ (Fig.~\ref{i:cdf2}).

\section{The case of a uniform EDF}
\label{appendix2}

\begin{figure}
\begin{center}$
\begin{array}{cc} 
\includegraphics[width=1\hsize]{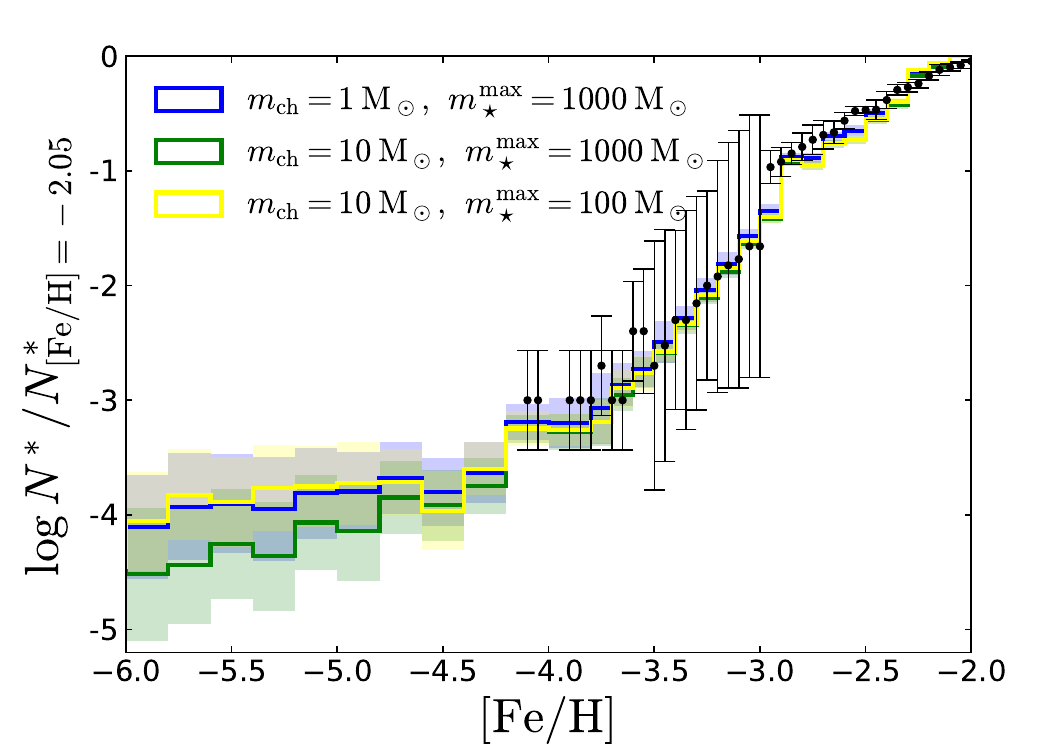} \\
\includegraphics[width=1\hsize]{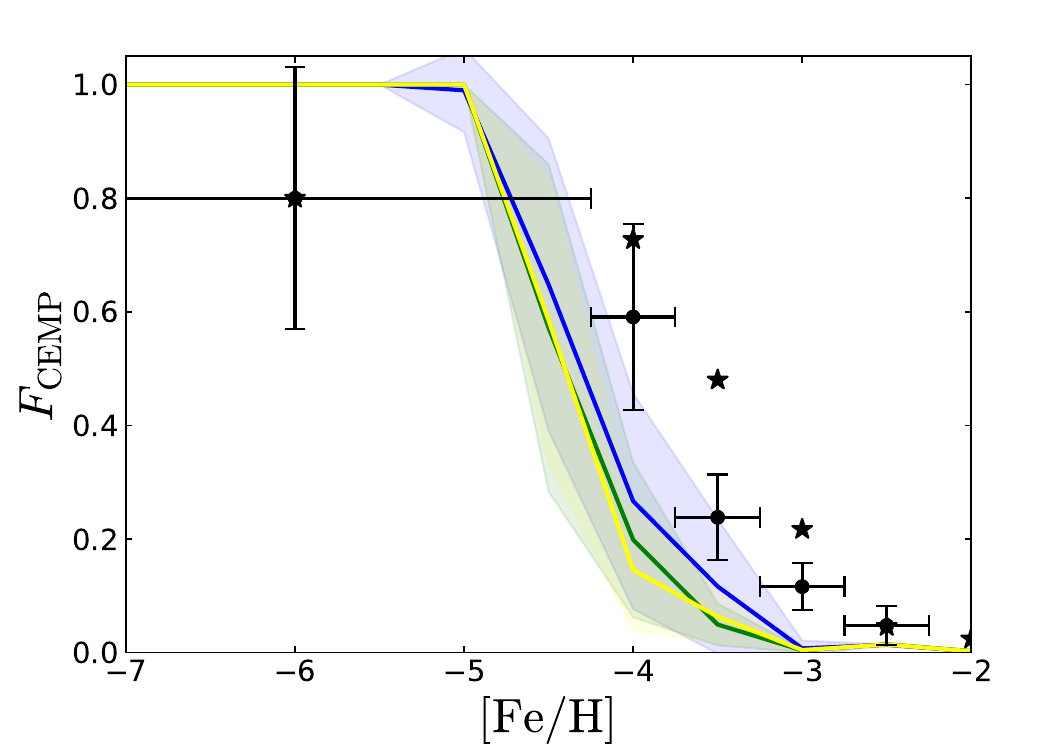}
\end{array}$
\end{center}
\caption{Metallicity distribution function (top) and CEMP fraction (bottom) of inner-halo stars when assuming that all Pop III stars with $m_\star=10-100\:{\rm M_\odot}$ have equal probability to explode as faint SNe, ccSNe, high energy SNe and hypernovae. The colors denote the characterictic mass, $m_{\rm ch}$ and the maximum mass, $m_\star^{\rm max}$ of the Pop III IMF considered in each model, as indicated in the legend. Datapoints show the observations of \citet{Bonifacio2021}, \citet{Yong2013a} and \citet{placco2014carbon} as in Fig.~\ref{i:mix}.}
\label{i:alleq}
\end{figure}

\begin{figure}
\begin{center}
\includegraphics[width=0.75\hsize]{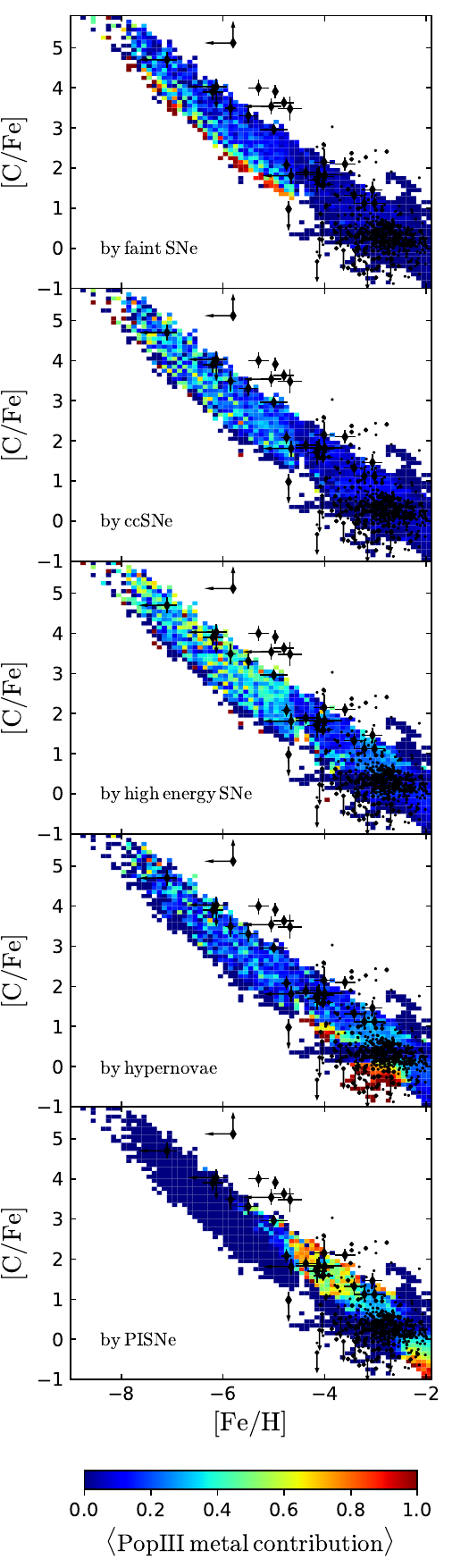} 
\end{center}
\caption{Same as Fig.~\ref{i:table} but for models in which Pop III stars with $m_\star=10-100\:{\rm M_\odot}$ have equal probability to explode as faint SNe, ccSNe, high energy SNe and hypernovae. }
\label{i:PLEGMAalleq}
\end{figure}

This appendix presents the results of models in which all types of primordial SNe (faint, cc, high energy SNe and hypernovae) with $m_\star=10-100\:{\rm M_\odot}$ are assumed to be equally probable. Three different Larson-type IMFs for Pop III stars are considered: two with a mass range $m_\star=0.1-1000\:{\rm M_\odot}$ and a characteristic mass $m_{\rm ch}=1\:{\rm M_\odot}$ and $m_{\rm ch}=10\:{\rm M_\odot}$ and one with $m_{\rm ch}=10\:{\rm M_\odot}$ and $m_\star=0.1-100\:{\rm M_\odot}$, i.e., that does not allow the formation of PISNe. The predicted MDFs and CEMP fractions in each case, are shown in Fig.~\ref{i:alleq}.  Fig.~\ref{i:PLEGMAalleq} shows the mean metal contribution from the different types of primordial SNe in each [C/Fe]--[Fe/H] bin.

The primordial EDF adopted here predicts a significantly lower CEMP-no fraction than the observations of \citet{Yong2013a} and \citet{placco2014carbon}. The descendants of the different Pop III SN types show similar properties as the ones in the models adopted in Section~\ref{Results}; faint SNe descendants occupy the [Fe/H]-poorest regions of the [C/Fe]--[Fe/H] diagram, while higher energy SNe dominate the metal enrichment at progressively higher [Fe/H]. Stars that have $\rm[C/Fe]<0$ are predominantly enriched by primordial hypernovae (at $\rm[Fe/H]\lesssim-2.5$) and PISNe (at $\rm [Fe/H]\gtrsim-2.5$). 

\bsp	
\label{lastpage}
\end{document}